\documentclass[journal=cmatex,manuscript=article]{achemso}

\usepackage{graphicx} 
\usepackage{chemformula}
\usepackage{siunitx}
\usepackage{chemfig}
\usepackage{caption}
\usepackage{subcaption}
\usepackage{hyperref}
\usepackage{wrapfig}
\usepackage{gensymb}

\title{Crystal Structures and Phase Stability of the Li$_2$S-P$_2$S$_5$ System from First Principles}
\author{Ronald L. Kam}
\affiliation{Materials Science Division, Lawrence Berkeley National Laboratory}
\alsoaffiliation{Department of Materials Science and Engineering, University of California, Berkeley}
\author{KyuJung Jun}
\affiliation{Materials Science Division, Lawrence Berkeley National Laboratory}
\alsoaffiliation{Department of Materials Science and Engineering, University of California, Berkeley}
\author{Luis Barroso-Luque}
\affiliation{Materials Science Division, Lawrence Berkeley National Laboratory}
\author{Julia H. Yang}
\affiliation{Materials Science Division, Lawrence Berkeley National Laboratory}
\alsoaffiliation{John A. Paulson School of Engineering and Applied Science, Harvard University (current affiliation)}
\author{Fengyu Xie}
\affiliation{Materials Science Division, Lawrence Berkeley National Laboratory}
\alsoaffiliation{Department of Materials Science and Engineering, University of California, Berkeley}
\author{Gerbrand Ceder}
\affiliation{Materials Science Division, Lawrence Berkeley National Laboratory}
\alsoaffiliation{Department of Materials Science and Engineering, University of California, Berkeley}
\email{gceder@berkeley.edu}
\date{February 2023}

\begin{document}

\maketitle
\section{Abstract}
The \ch{Li2S}-\ch{P2S5} pseudo-binary system has been a valuable source of promising superionic conductors, with $\alpha$-\ch{Li3PS4}, $\beta$-\ch{Li3PS4}, HT-\ch{Li7PS6}, and \ch{Li7P3S11} having excellent room temperature \ch{Li}-ion conductivity $>$ 0.1 mS/cm. The metastability of these phases at ambient temperature motivates a study to quantify thermodynamic accessibility. Through calculating the electronic, configurational, and vibrational sources of free energy from first principles, a phase diagram of the crystalline \ch{Li2S}-\ch{P2S5} space is constructed. Well-established phase stability trends from experiments are recovered, such as polymorphic phase transitions in \ch{Li7PS6} and \ch{Li3PS4}, and the metastability of \ch{Li7P3S11} at high temperature. At ambient temperature, it is predicted that all superionic conductors in this space are indeed metastable, but thermodynamically accessible. Vibrational and configurational sources of entropy are shown to be essential towards describing the stability of superionic conductors. New details of the Li sublattices are revealed, and are found to be crucial towards accurately predicting configurational entropy. All superionic conductors contain significant configurational entropy, which suggests an inherent correlation between superionic conductivity and high configurational entropy.

\section{Introduction}
\begin{figure}[t!]
    \centering
    \begin{subfigure}[b]{0.95\linewidth}
        \centering
        \includegraphics[scale=0.45]{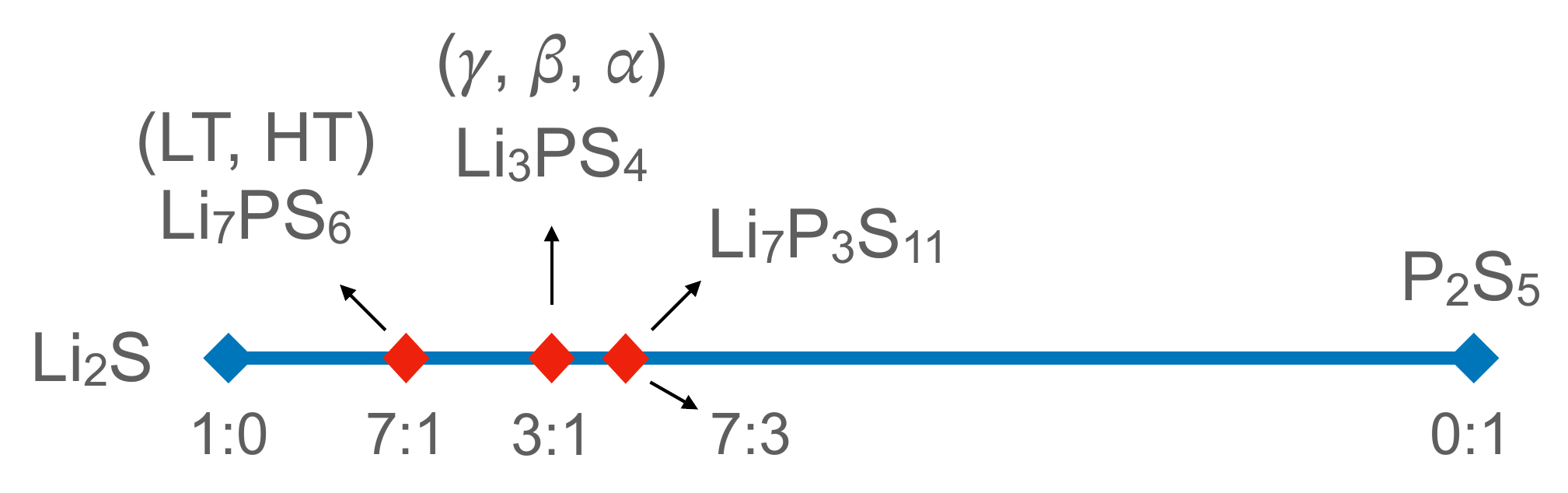}
        \caption{Composition line}
        \label{fig:lps_comp_line}
    
    \end{subfigure}
    \hspace{0.2cm}
    \begin{subfigure}[b]{0.3\linewidth}
        \centering
        \includegraphics[scale=0.25]{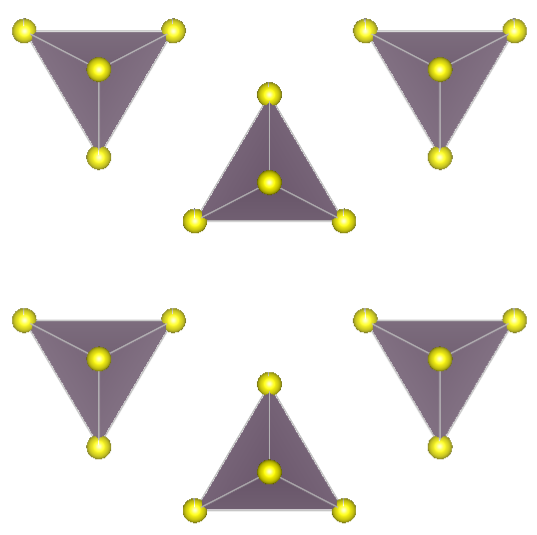}
        \caption{$\gamma$-\ch{Li3PS4}}
        \label{fig:gamma_ps4}
    \end{subfigure}
    \hspace{0.2cm}
    \begin{subfigure}[b]{0.3\linewidth}
        \centering
        \includegraphics[scale=0.28]{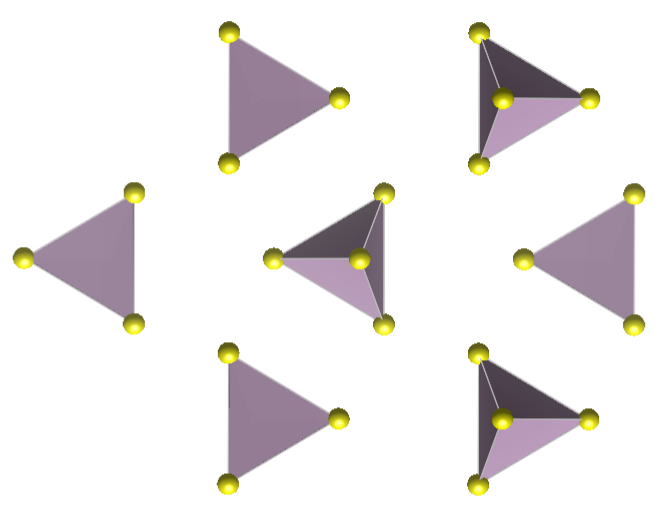}
        \caption{$\beta$-\ch{Li3PS4}}
        \label{fig:beta_ps4}
    \end{subfigure}
    \hspace{.2cm}
    \begin{subfigure}[b]{0.3\linewidth}
        \centering
        \includegraphics[scale=0.3]{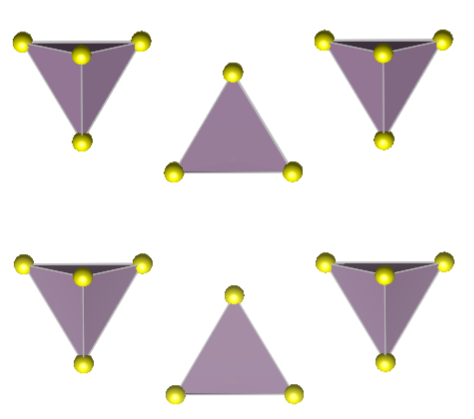}
        \caption{$\alpha$-\ch{Li3PS4}}
        \label{fig:alpha_ps4}
    \end{subfigure}
    \hspace{0.2cm}
    
    \begin{subfigure}[b]{0.3\linewidth}
        \centering
        \includegraphics[scale=0.3]{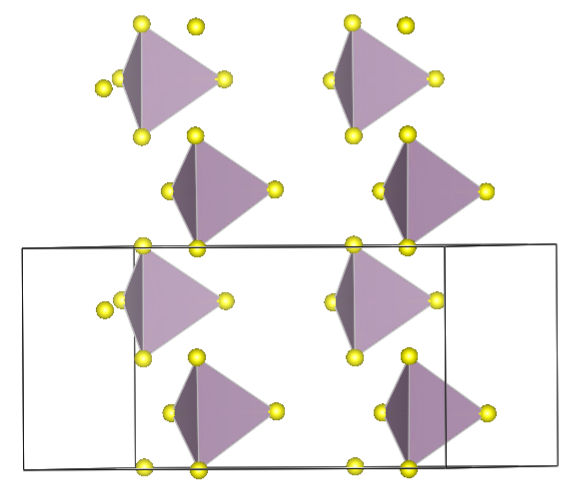}
        \caption{LT-\ch{Li7PS6}}
        \label{fig:lt_ps4}
    \end{subfigure}
    \hspace{0.2cm}
    \begin{subfigure}[b]{0.3\linewidth}
        \centering
        \includegraphics[scale=0.28]{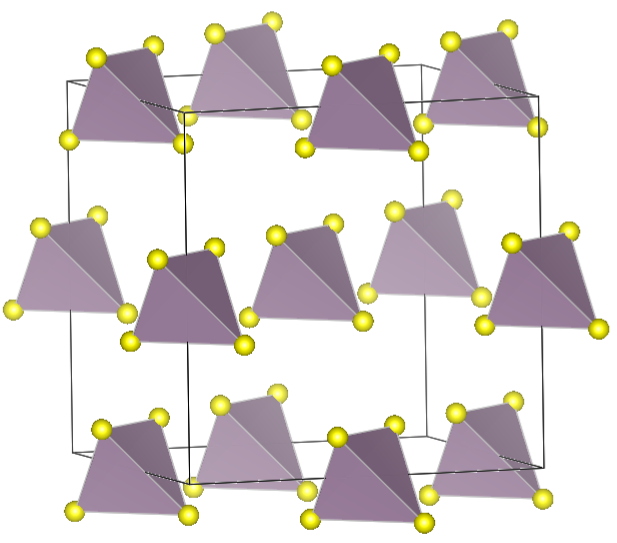}
        \caption{HT-\ch{Li7PS6}}
        \label{fig:ht_ps4}
    \end{subfigure}
    \hspace{0.2cm}
    \begin{subfigure}[b]{0.3\linewidth}
        \centering
        \includegraphics[scale=0.26]{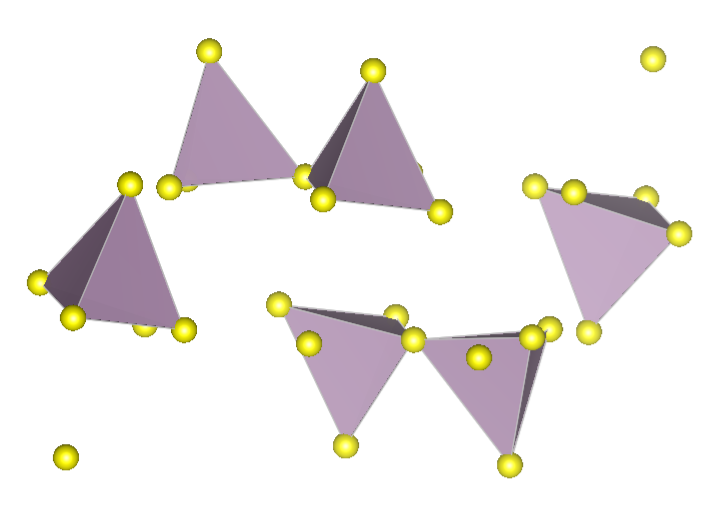}
        \caption{\ch{Li7P3S11}}
        \label{fig:7311_ps4_p2s7}
    \end{subfigure}
    \label{comp_and_ps4}
    \caption{Composition line and arrangements of \ch{PS4} tetrahedral units in the \ch{Li2S}-\ch{P2S5} system. a) Location of \ch{Li7PS6}, \ch{Li3PS4}, and \ch{Li7P3S11} on the composition line, labeled by the ratios of \ch{Li2S} to \ch{P2S5}. Arrangements of \ch{PS4} and \ch{P2S7} units in a) $\gamma$-\ch{Li3PS4}, b) $\beta$-\ch{Li3PS4}, c) $\alpha$-\ch{Li3PS4}, d) HT-\ch{Li7PS6}, e) LT-\ch{Li7PS6}, f) \ch{Li7P3S11}. Unit cell boxes are drawn for LT-\ch{Li7PS6} and HT-\ch{Li7PS6} to show their cubic and orthorhombic \ch{PS4} arrangement, respectively.}
\end{figure}

The global transition to sustainable energy sources necessitates the continued development of energy storage technologies that enable increased deployment of intermittent energy sources (ie wind and solar power) and electrification of transportation \cite{goodenough_perspective_2013}. Lithium (\ch{Li}) all solid-state batteries (ASSB) can significantly improve the safety and energy density compared to conventional \ch{Li}-ion batteries using organic liquid electrolytes \cite{janek_zeier_2016_nature_energy, lee_ssb_agc_2020, kato_high-power_2016}. Discovery and development of novel superionic conductors with Li-ion conductivity on the order of organic liquid electrolytes ($>0.1$ mS/cm) is crucial towards enabling ASSBs to have similar power densities as conventional Li-ion batteries \cite{kato_high-power_2016}. The pseudo-binary \ch{Li2S}-\ch{P2S5} composition space has proven to be a particularly rich source of promising Li superionic conductors. Several crystalline compounds can be synthesized by combining \ch{Li2S} and \ch{P2S5} precursors in varying ratios (Figure \ref{fig:lps_comp_line})\cite{kudu_li2s_p2s5_review_2018}, with the notable phases being the $\alpha$, $\beta$, and $\gamma$ polymorphs of \ch{Li3PS4}, high-temperature (HT) and low-temperature (LT)-\ch{Li7PS6}, and \ch{Li7P3S11}. Among these, $\alpha$-\ch{Li3PS4}, $\beta$-\ch{Li3PS4}, HT-\ch{Li7PS6}, and \ch{Li7P3S11} are superionic conductors \cite{ziolkowska_li7ps6_liquid_2019, seino_sulphide_2014}. Although amorphous phases with these compositions also exist\cite{byungju_paddle_2023, mizuno_lps_glasses_2006, guo_ai_lps_pd}, the focus of our study will be on understanding the relative phase stability of the crystalline phases only.

The crystalline phases in the \ch{Li2S}-\ch{P2S5} space are composed of periodically arranged \ch{PS4} tetrahedra, which are either isolated or form \ch{P2S7} ditetrahedra. Li atoms are located in between these units and coordinated by S atoms. The \ch{Li7PS6} polymorphs also contain free S atoms that are only coordinated with Li atoms. Different phases can be identified by their distinct orientation of \ch{PS4} and \ch{P2S7} groups, which are shown in Figure 1. The \ch{Li3PS4} and \ch{Li7PS6} polymorphs are all composed of isolated \ch{PS4} groups. In $\gamma$-\ch{Li3PS4}, these \ch{PS4} groups are uni-directional, with all apexes facing the same direction (apexes face out of the page in Figure \ref{fig:gamma_ps4}). In $\beta$-\ch{Li3PS4}, \ch{PS4} groups are arranged in alternating zig-zag chains, with each chain containing apexes that face the same direction, while apexes in the adjacent chain face the opposite direction (Figure \ref{fig:beta_ps4}) \cite{homma_li3ps4_crystal_2011}. The $\alpha$-\ch{Li3PS4} polymorph also contains \ch{PS4} with oppositely facing apexes, but these are arranged in alternating columns (Figure \ref{fig:alpha_ps4}). In both \ch{Li7PS6} polymorphs, all \ch{PS4} groups face the same direction, but differ in their spatial distributions \cite{kong_2020_li7ps6_xrd}. In LT-\ch{Li7PS6}, \ch{PS4} are arranged with orthorhombic symmetry (Figure \ref{fig:lt_ps4}) , while in HT-\ch{Li7PS6} the \ch{PS4} are arranged with face centered cubic (FCC) symmetry (Figure \ref{fig:ht_ps4}) \cite{kong_2020_li7ps6_xrd}. \ch{Li7P3S11} is composed of both \ch{P2S7} and \ch{PS4} units (Figure \ref{fig:7311_ps4_p2s7})\cite{yamane_li7p3s11_xrd_2007}.

According to previous experimental and computational studies, the superionic conductor phases are all metastable at ambient temperature \cite{kudu_li2s_p2s5_review_2018, guo_ai_lps_pd}. Among the \ch{Li3PS4} polymorphs, $\gamma$ is the stable phase at room temperature but has low Li conductivity, while $\beta$ and $\alpha$ are the high temperature fast-conducting phases \cite{homma_li3ps4_crystal_2011, kaup_li3ps4_nd_2020}. $\beta$-\ch{Li3PS4} has been stabilized at room temperature as nanoporous particles from solution-state synthesis \cite{liu_anomalous_2013, makiura_surf_stab_sic}. This phase has also been stabilized through mechanochemical synthesis involving ball milling to form an amorphous phase, and a subsequent heat treatment to recrystallize \cite{mizuno_lps_glasses_2006}. An analogous Si-doped Li$_{3.25}$Si$_{0.25}$P$_{3.75}$S$_4$ structure, where Si substitutes into phosphorus (P) sites, has also been stabilized at room temperature \cite{zhou_entropically_2019}. $\alpha$-\ch{Li3PS4} has recently also been stabilized at room temperature via a rapid heating and quenching technique \cite{alpha_synthesis_2023}. This discovery indicates that the energy differences between the three \ch{Li3PS4} polymorphs at room temperature should be small, allowing for the metastable $\alpha$ and $\beta$ to be thermodynamically accessible at ambient temperature. HT-\ch{Li7PS6} is only stable at elevated temperatures (T $>$ 483 K) \cite{kong_2020_li7ps6_xrd}, but has been successfully stabilized at room temperature through halide atom substitution into S sites, typically to form the \ch{Li6PS5X} composition (X = Cl, Br, or I) \cite{schlenker_li6ps5cl_nd_pdf, kong_li7ps6_se_nmr, adeli_argyrodite_subs_2019, kraft_polariz_li6ps5x}. Synthesis of \ch{Li7P3S11} usually requires ball-milling to its amorphous form before recrystallization above its glass transition temperature of around 500 K \cite{mizuno_li7p3s11_crystal_2005, yamane_li7p3s11_xrd_2007, kudu_li2s_p2s5_review_2018}. Heat treatment at higher temperatures (T $>$ 800 K) is not possible, as \ch{Li7P3S11} phase-separates to \ch{Li4P2S6} and \ch{Li3PS4} \cite{mizuno_li7p3s11_crystal_2005}.

The metastable nature of these superionic conductors motivates our first-principles study with the objective to understand their thermodynamic accessibility at finite temperature, rationalize experimental trends, and potentially propose new synthesis procedures. To model the free energy of each phase, we consider contributions from the electronic structure, configurational disorder, and vibrational modes.  We find that including both configurational and vibrational entropy is necessary to correctly predict free energies, in agreement with a previous study on the Li$_{1+2x}$Zn$_{1+x}$PS$_4$ system \cite{richards_design_2016}.

We model configurational Li-vacancy disorder with well-established lattice model methods \cite{barroso_luque_ionic_ce_2022, gerd_ising_1993}, which have been previously used to study a range of alkali-ion intercalation oxides and solid electrolytes \cite{vdv_lco_phase_1998, vdv_alkali_theory_2020, deng_nasicon_ce_2020}. To properly model the configurational disorder in HT-\ch{Li7PS6}, $\alpha$-\ch{Li3PS4}, $\beta$-\ch{Li3PS4}, and \ch{Li7P3S11}, we require accurate structural models to define the set of distinct sites that Li can occupy, which we refer to as the Li sublattice. There are conflicting reports about the specific sites that make up the Li sublattices arising from different characterization techniques. More specifically, in $\alpha$-\ch{Li3PS4}, $\beta$-\ch{Li3PS4}, and HT-\ch{Li7PS6}, neutron diffraction (ND) refinements \cite{kaup_li3ps4_nd_2020, schlenker_li6ps5cl_nd_pdf} have identified more Li sites and increased site disorder as compared to X-ray diffraction (XRD) refinements 
\cite{homma_li3ps4_crystal_2011, kanno_alpha_li3ps4_structure, kong_2020_li7ps6_xrd}. In \ch{Li7P3S11}, XRD and ND identify fully ordered, but entirely different Li sublattices \cite{yamane_li7p3s11_xrd_2007, onodera_li7p3s11_nd_xrd}. A more recent ab-initio molecular dynamics (AIMD) study proposing 15 potential Li sites in \ch{Li7P3S11} introduces uncertainty to the exact state of Li order, since these new sites can in principle be partially occupied \cite{chang_7311_aimd}. Because of these conflicting reports, we dedicate a large portion of this study towards clarifying the Li arrangement in these structures, the details of which we find to be essential for recovering experimental thermodynamic trends.

For each disordered phase, we assess the validity of various proposed Li sublattices, primarily by analyzing atomic relaxation distances and comparing Li site disordering behavior to experimental reports. Upon obtaining the most representative Li sublattice, we train a cluster expansion (CE), which can rapidly evaluate total energies of any Li-vacancy configuration within the given Li sublattice \cite{barroso_luque_ionic_ce_2022}. Using the CE, we perform Monte Carlo (MC) sampling to determine the configurational entropy, free energy, and Li site disordering behavior as a function of temperature.

The CE formally represents the energy of a disordered crystal structure as a summation over contributions from local, multi-site (cluster) configurations and their associated interaction energies\cite{sanchez_generalized_cluster_1984, barroso_luque_ionic_ce_2022}. The expression for CE energy is shown in equation \ref{ce_energy}, where $\Vec{\sigma}$ is the vector encoding the species occupying each lattice site, $\beta$ is the index for a symmetrically distinct cluster, $J_{\beta}$ is the effective cluster interaction (ECI) energy, and $\langle\Phi(\sigma)\rangle_\beta$ is the correlation function describing the crystal-averaged cluster configuration. The ECI are determined from regularized linear regression techniques, using a training set of distinct DFT-relaxed configurations and energies \cite{barroso_luque_ionic_ce_2022}. \begin{equation}\label{ce_energy}
    E(\Vec{\sigma}) = \sum_\beta J_\beta \langle\Phi(\Vec{\sigma})\rangle_\beta
\end{equation}
Monte Carlo (MC) sampling is then performed using the CE in the canonical ensemble to predict Li site disorder, identify new ground state (lowest energy) structures, and calculate configurational thermodynamic properties through thermodynamic integration (more in Methods).

Vibrational free energy contributions are captured in the ground state of each phase, by performing harmonic phonon calculations \cite{togo_phonopy_2015}. By incorporating the contributions to the free energy from electronic structure, vibrational entropy, and configurational entropy, we assess thermodynamic stability in the \ch{Li2S}-\ch{P2S5} phase space, recovering well-established experimental observations.

This paper is organized as follows. 1) We first present the pseudo-binary \ch{Li2S}-\ch{P2S5} phase diagram. The thermodynamic stability of each phase at finite temperature is evaluated and potential synthesis procedures for metastable phases are proposed. 2) For each composition, we discuss the appropriate choice of refined structure for each polymorph by comparing the validity of previously proposed models. Phase stability trends between polymorphs are examined in detail, with a focus on identifying phase transitions and quantifying the contributions of vibrational and configurational entropy towards stability. 3) In the discussion, we draw further connections to previously proposed experimental synthesis strategies, and explore a potential correlation between superionic conductivity and high configurational entropy.

\section{Results}

\subsection{Phase stability in the \ch{Li2S}-\ch{P2S5} system}
\begin{figure}[t!]
    \centering
    \begin{subfigure}[b]{0.98\linewidth}
        \centering
        \includegraphics[scale=0.85]{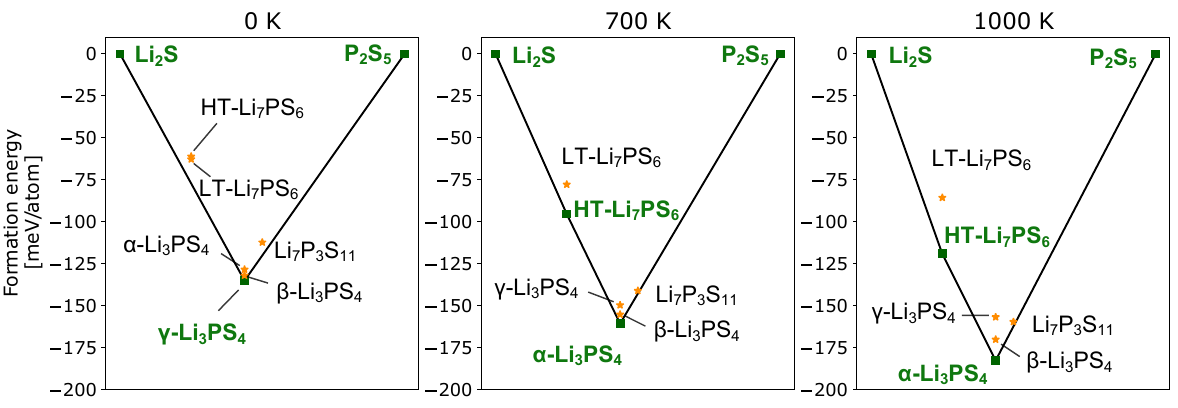}
        \caption{}
        \label{fig:binary_pd}
    \end{subfigure}
    \begin{subfigure}[b]{0.95\textwidth}
        \centering
        \includegraphics[scale=0.65]{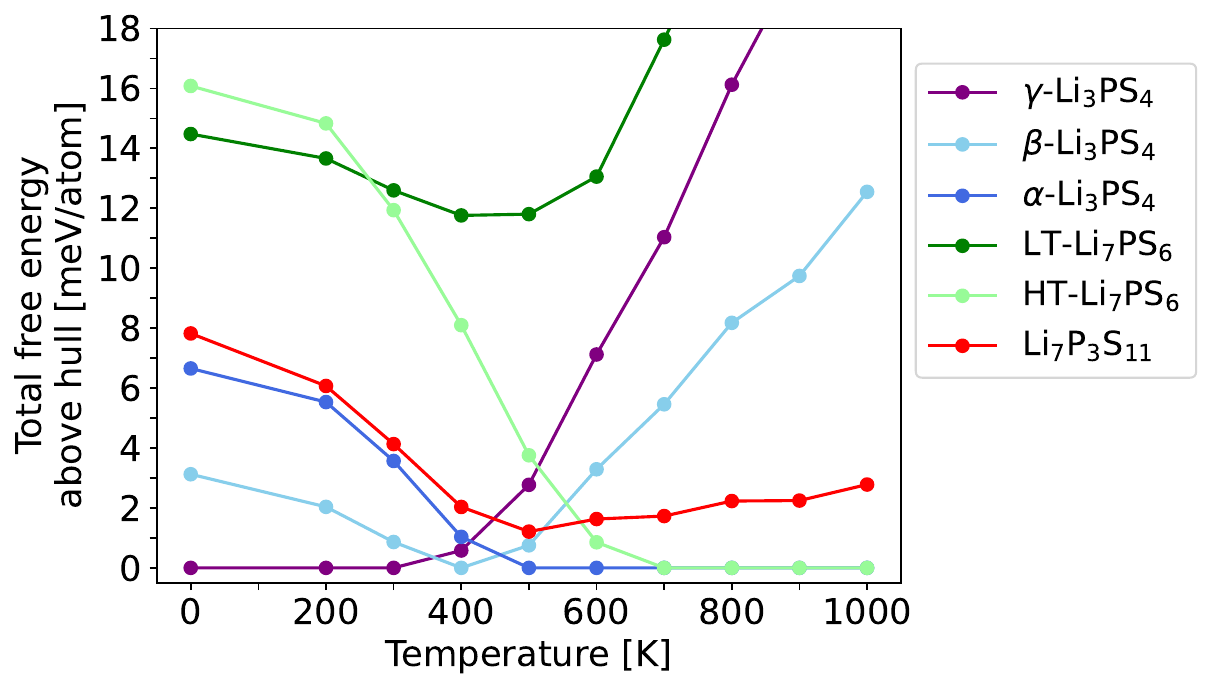}
        \caption{}
        \label{fig:e_hull}
    \end{subfigure}
    
    \caption{Phase stability of the \ch{Li2S}-\ch{P2S5} pseudo-binary system. a) Phase diagram at 0 K, 700 K, and 1000 K. Solid lines denote the convex hull. Stable phases that are on the convex hull are marked with squares and labeled in green. Unstable phases are marked with gold stars and labeled in black. b) Free energy above hull for all phases from 0 to 1000 K.}
    \label{fig:phase-stability}
\end{figure}
The pseudo-binary \ch{Li2S}-\ch{P2S5} phase diagram is presented in Figure \ref{fig:binary_pd} and the energies above the hull (E$_\text{hull}$) as a function of temperature are shown in Figure \ref{fig:e_hull}. The convex hull is a typical construction to obtain stable phases and represents the collection of thermodynamic ground states into which all other phases have a driving force to convert. All computed formation free energies used to construct the phase diagram are shown in SI Figure S7. At 0 K, the only stable phases on the convex hull are $\gamma$-\ch{Li3PS4} and the endpoints, \ch{Li2S} and \ch{P2S5} (Figure \ref{fig:binary_pd}). At 700 K, HT-\ch{Li7PS6} is stabilized  and appears on the hull. Since reported synthesis procedures for LT and HT-\ch{Li7PS6} typically do not require mechanical milling or quenching \cite{kong_2020_li7ps6_xrd, kong_li7ps6_se_nmr}, it may be surprising that they are unstable at 300 K---13 and 12 meV/atom above the hull respectively (Figure \ref{fig:e_hull}). It is likely that the thermodynamically favored phase separation of HT-\ch{Li7PS6} to \ch{Li2S} and \ch{Li3PS4} is kinetically hindered at room temperature. Instead, HT-\ch{Li7PS6} is found to transform to LT-\ch{Li7PS6} upon cooling, a potentially more facile process as it merely involves shifting the \ch{PS4} locations (Figure 1). Thus, an appropriate solid-state synthesis procedure would be to perform sufficiently high temperature (T $> 600$ K) synthesis to stabilize HT-\ch{Li7PS6}, before a relatively rapid cooling process to bypass the phase separation to \ch{Li2S} and \ch{Li3PS4}.

For the \ch{Li3PS4} composition, our calculations in Figure \ref{fig:e_hull} predict phase transformations from $\gamma$ $\rightarrow$ $\beta$ $\rightarrow$ $\alpha$ with increasing temperature, which is consistent with experiments. Since $\beta$-\ch{Li3PS4} is less than 1 meV/atom above the hull at 300 K (Figure \ref{fig:e_hull}), it is plausible that nanoporous synthesis and mechanical milling techniques can lead to its stabilization at room temperature \cite{liu_anomalous_2013, mizuno_lps_glasses_2006}. The $\alpha$-\ch{Li3PS4} polymorph is only slightly less stable than $\beta$ at 300 K (E$_{\text{hull}}$ = 4 meV/atom), which explains why $\alpha$ can also be stabilized at ambient temperature through a rapid heating and quenching procedure \cite{alpha_synthesis_2023}. Rapid heating of the \ch{Li3PS4} glass to temperatures in the stability range of $\beta$ enables nucleation of metastable $\alpha$ particles that are only slightly less stable than $\beta$, which is possible by the Ostwald step rule \cite{alpha_synthesis_2023, ostwald_rule}. Rapid quenching can then obstruct the commonly observed direct transition from $\alpha$ to $\gamma$ \cite{homma_li3ps4_crystal_2011, kaup_li3ps4_nd_2020}, which is possible as their energy difference is only 4 meV/atom at 300 K.
 
\ch{Li7P3S11} (red curve in Figure \ref{fig:e_hull}) is metastable across all temperatures as its energy is never low enough to be on the convex hull, which agrees with prior experimental studies \cite{mizuno_li7p3s11_crystal_2005, yamane_li7p3s11_xrd_2007}. At 300 K, it is 4 meV/atom above the convex hull. As temperature increases to 500 K, its E$_{\text{hull}}$ decreases to a minimum of 1.4 meV/atom. Further increases in temperature lead to greater E$_{\text{hull}}$. Thus, an ideal synthesis temperature should be around 500 K, corresponding to the minimum E$_{\text{hull}}$. This temperature is remarkably close to its experimentally observed glass transition temperature and helps rationalize why heat treatments near this temperature have been successful for recrystallization \cite{mizuno_li7p3s11_crystal_2005, seino_sulphide_2014}. The increasing instability with respect to temperature helps explain the experimentally observed tendency to phase separate to \ch{Li3PS4} and \ch{Li4P2S6} at temperatures greater than 800 K \cite{mizuno_lps_glasses_2006}. The source of this instability is the competition with $\alpha$-\ch{Li3PS4}, its neighboring stable point, which lowers its free energy more with increasing temperature, therefore increasing the convex hull depth (Figure \ref{fig:binary_pd}). We will show in the next section that this arises from the high configurational entropy in $\alpha$-\ch{Li3PS4}.

\subsection{\ch{Li3PS4} polymorphs}
$\gamma$-\ch{Li3PS4} (Pnm2$_1$) is the stable polymorph at room temperature and has a very low Li conductivity of $3 (10^{-4})$ mS/cm \cite{homma_li3ps4_crystal_2011}. The reported XRD and ND refinements are in excellent agreement with each other, showing an ordered Li sublattice comprising of fully occupied Li1 (4b) and Li2 (2a) sites. Since there is no ambiguity in these refinements, we use this structure to model $\gamma$-\ch{Li3PS4} \cite{homma_li3ps4_crystal_2011, kaup_li3ps4_nd_2020}.

\subsubsection{$\beta$-\ch{Li3PS4} structure}

Upon heating, $\gamma$ transforms to $\beta$-\ch{Li3PS4} at around 575 K, crystallizing in the orthorhombic Pnma space group, which leads to a lattice volume expansion by $\sim$3$\%$ \cite{homma_li3ps4_crystal_2011}. The zig-zag ordering of \ch{PS4} units generates a different Li sublattice with more sites than in $\gamma$, leading to the potential for disorder. At around 600 K, XRD refinements have reported Li atoms occupying Li1' (8d), Li2' (4b), and Li3' (4c) sites, with fractional occupancies of 1, 0.7, and 0.3, respectively (XRD refined sites are labeled with apostrophes and ND refined sites without apostrophes for clarity in this discussion)\cite{homma_li3ps4_crystal_2011}. A more recent ND refinement proposes a slightly different model, with reported site splitting of Li1’ (8d) to Li1A (8d) and Li1B (8d), and splitting of Li2' (4b) to Li2 (8d), while retaining its Pnma symmetry \cite{kaup_li3ps4_nd_2020}. The 4 distinct Li sites refined by ND are all partially occupied.

\begin{figure}[t!]
\centering
    \begin{subfigure}[b]{0.95\linewidth}
        \centering
        \includegraphics[scale=0.4]{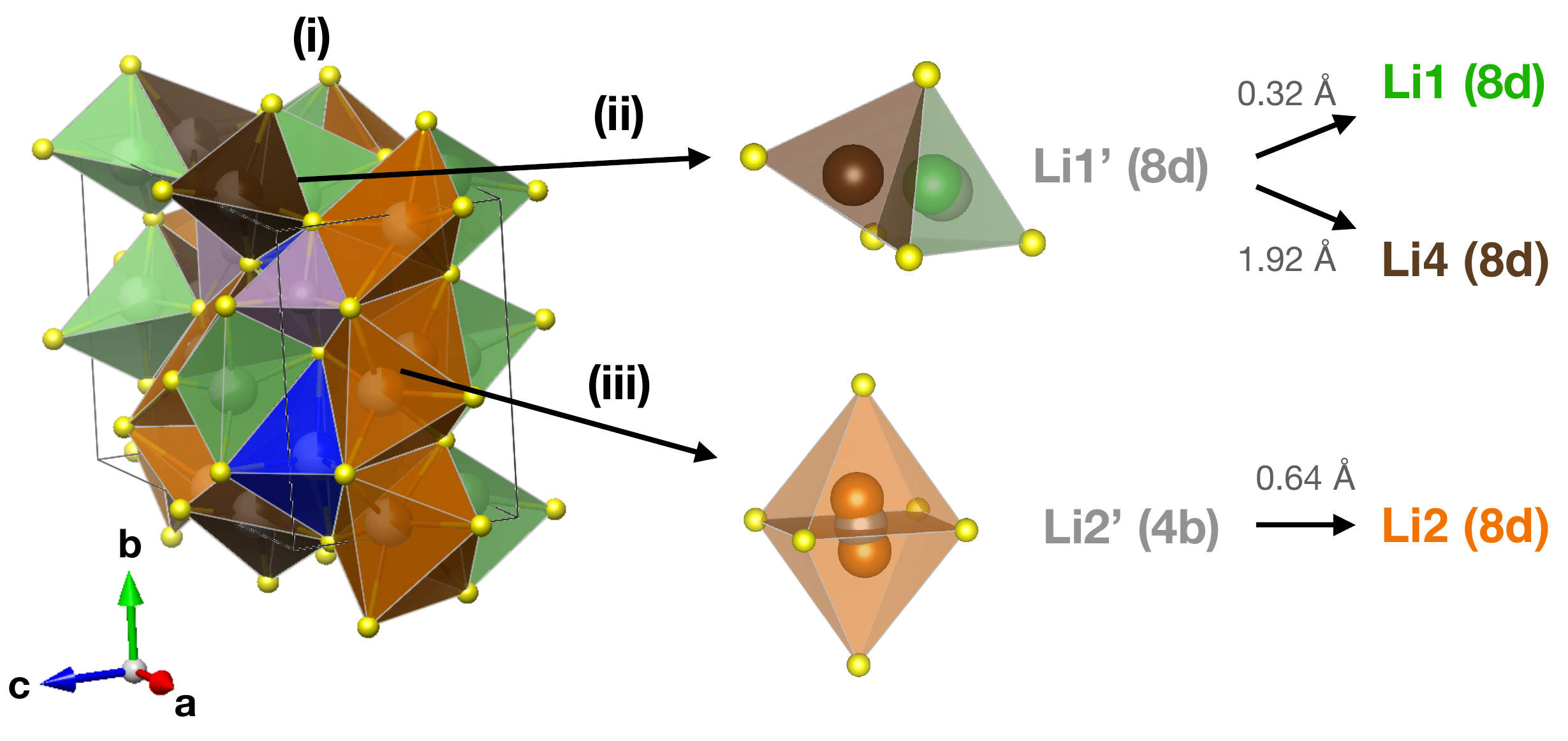}
        \caption{}
        \label{fig:b_lps_unit_splitting}
    \end{subfigure}
    \hspace{0.3cm}
    \begin{subfigure}[b]{0.95\linewidth}
        \centering
        \includegraphics[scale=0.6]{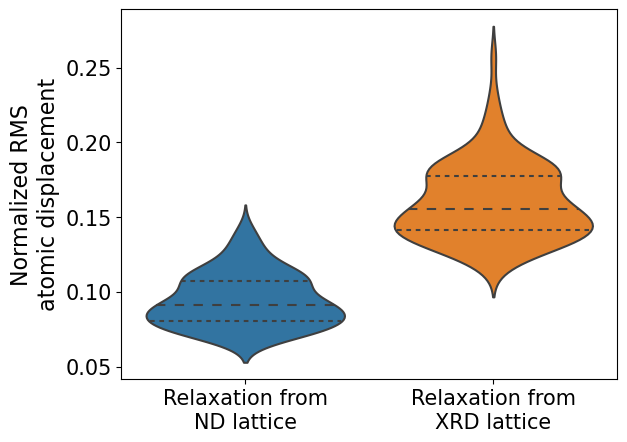}
        \caption{}
        \label{fig:li3ps4_site_relax}
    \end{subfigure}
\caption{Structure of $\beta$-\ch{Li3PS4}. a) (i) Unit cell with Li1 (8d) (green), Li2 (8d) (orange), Li3 (4c) (blue), and Li 4 (8d) (brown) sites. (ii) Splitting of Li1' (8d) (grey) to Li1 (8d) and Li4 (8d). (iii) Splitting of Li2' (4b) (grey) from square planar coordination to 5-fold coordinated Li2 (8d). b) Distributions of atomic relaxations in DFT starting from ideal ND and XRD refined structures. The extent of atomic relaxations are measured in normalized root mean squared (NRMS) atomic displacements. The dashed lines denote the 25, 50, 75 percentiles in the distributions.}
\end{figure}

We analyze the geometric discrepancy between XRD and ND refinements of $\beta$-\ch{Li3PS4} by inspecting the Li coordination environments in the XRD and ND sites. In Figure \ref{fig:b_lps_unit_splitting}, we show (i) the unit cell, (ii) splitting of Li1' (8d), and (iii) splitting of Li2' (4b). The splitting of the Li1' (8d) site (grey in Figure \ref{fig:b_lps_unit_splitting}ii) in fact yields two distinct sites: Li1A (8d) and Li1B (8d) (green and brown in Figure \ref{fig:b_lps_unit_splitting}ii, respectively). Li1A (8d) is essentially identical to Li1' (8d), while Li1B (8d) is its face sharing neighbor 1.7 $\AA$ away. The emergence of Li1B as a new Li site can be detected by ND, while in XRD it has not been detected, likely due to the small X-ray scattering factor of Li. Since Li1A and Li1B sites are not related to each other, we will refer to Li1A as Li1 (8d) and Li1B as Li4 (8d) in the following discussion. Li2' (4b) (grey in Figure \ref{fig:b_lps_unit_splitting}iii), with square planar coordination, splits into two neighboring and face-sharing Li2 (8d) sites (orange in Figure \ref{fig:b_lps_unit_splitting}iii), each with 5 fold coordination. XRD was unable to distinguish the two neighboring Li2 (8d) sites, which are only 1.3 $\AA$ apart, and instead identified just one Li2' (4b) site.

To assess the accuracy of XRD and ND refinements of $\beta$-\ch{Li3PS4}, we examine for all atomic positions in the DFT relaxed configurations the deviation from their XRD and ND refined sites. This is measured by calculating the normalized root mean squared (NRMS) displacement of relaxed atomic locations from the ND and XRD refined $\beta$-\ch{Li3PS4} lattices. The atoms of a relaxed structure are mapped back to a refined lattice site to construct the "refined" structure. The atoms of the relaxed and refined structures are then placed on an averaged lattice (in Cartesian coordinates) that minimizes the NRMS displacement, which is defined in equation \ref{nrms}, where $\Delta x_i$  is the displacement of atom i between the DFT-relaxed structure and ND or XRD-refined model in Cartesian coordinates, $N$ is the number of atoms, and V is the cell volume \cite{ong_pymatgen_2013}.
\begin{equation}\label{nrms}
    \text{NRMS displacement} = \frac{\sqrt{\sum_i ^N \Delta x_i^2 / N}} {(V/N)^{1/3}}
\end{equation}
In Figure \ref{fig:li3ps4_site_relax}, we show violin plots of the distributions of NRMS displacement from the ND-refined structure (blue) and XRD-refined structure (orange). NRMS displacement from the ND structure is significantly smaller compared to the XRD structure, since the 3rd quartile of the ND and the 2nd quartile of the XRD distributions do not overlap (Figure \ref{fig:li3ps4_site_relax}). The distribution of relaxations from the ND structure also has a smaller range, and thus less probable outliers, suggesting that the ND refinement is more accurate.

\begin{figure}
\centering
    \begin{subfigure}[b]{0.4\textwidth}
        \includegraphics[scale=0.3]{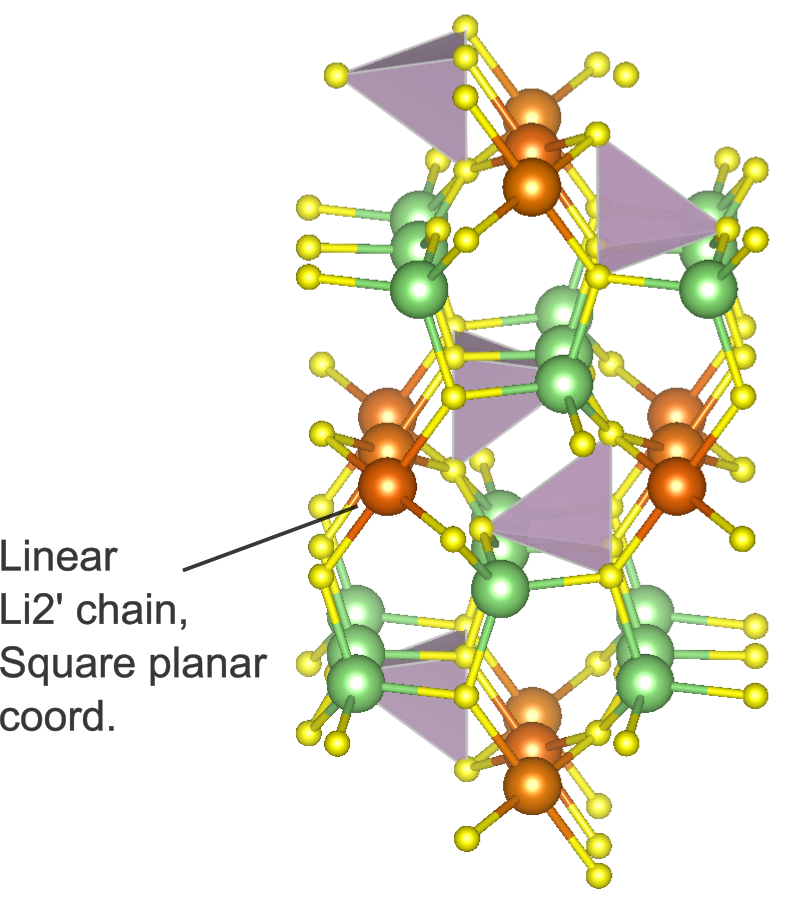}
        \caption{}
        \label{fig:lps_b_hsgs_struc}
    \end{subfigure}
    \hspace{0.05cm}
    \begin{subfigure}[b]{0.4\textwidth}

        \includegraphics[scale=0.3]{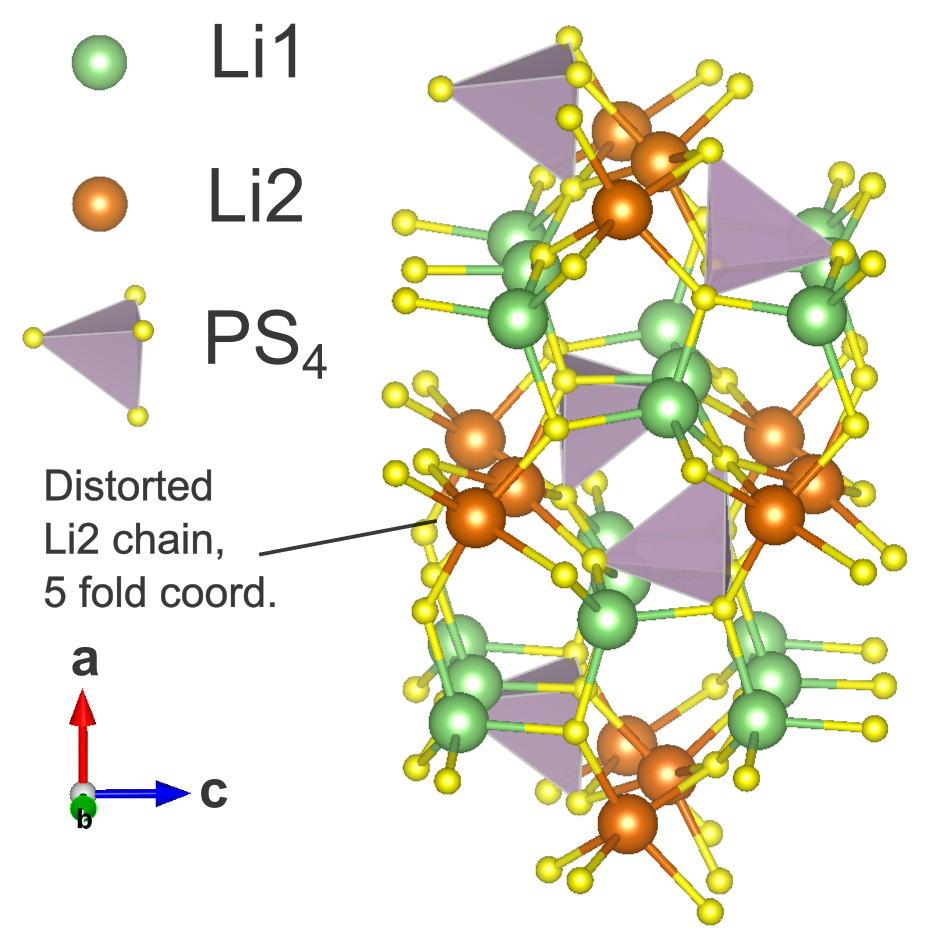}
        \caption{}
        \label{fig:lps_b_lsgs_struc}
        
    \end{subfigure}
    \begin{subfigure}[b]{0.45\textwidth}
        \includegraphics[scale=0.65]{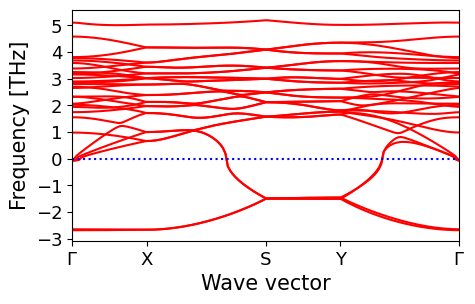}
        \caption{}
        \label{fig:lps_b_hsgs_band}
    \end{subfigure}
    \hfill
    \begin{subfigure}[b]{0.45\textwidth}
        \includegraphics[scale=0.65]{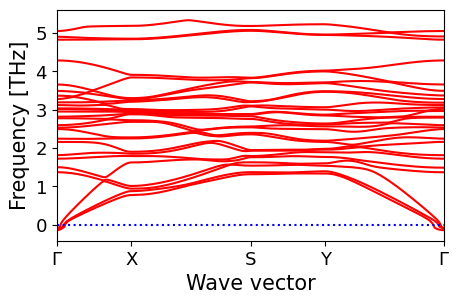}
        \caption{}
        
        \label{fig:lps_b_lsgs_band}
    \end{subfigure}
    \begin{subfigure}[b]{0.45\textwidth}
        \includegraphics[scale=0.3]{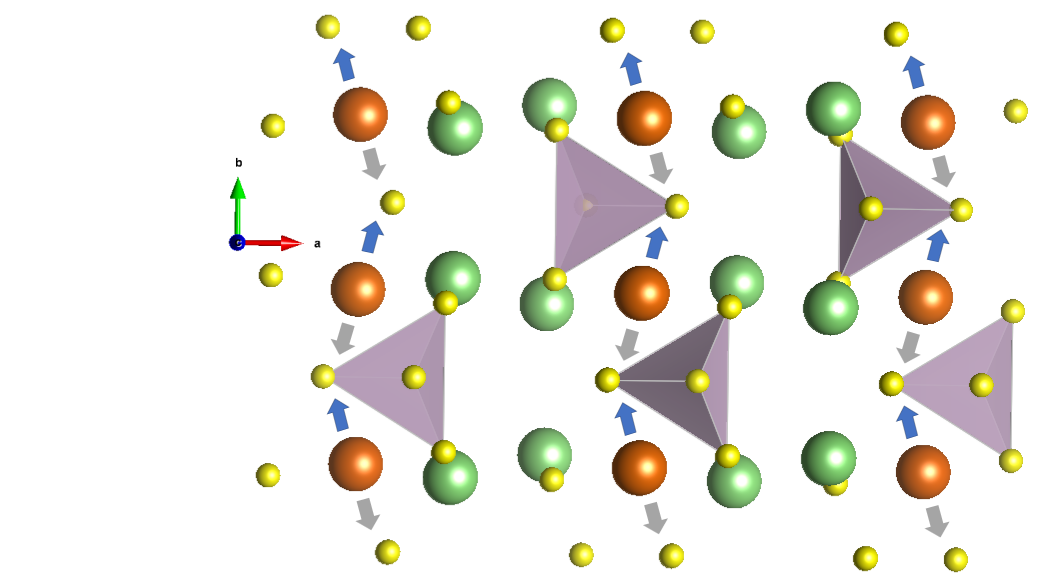}
        \caption{}
        \label{fig:lps_b_hsgs_mode}
    \end{subfigure}
\caption{Structures of the a) XRD ground state (XRD-GS) and b) ND ground state (ND-GS) of $\beta$-\ch{Li3PS4}. XRD-GS contains linear ordering of Li1 and Li2 atoms along [010] and [001], while the ND-GS has distorted, staggered chains along [010]. Phonon dispersions of the c) XRD-GS and (d) ND-GS. e) Visualizing an imaginary optical mode of XRD-GS at $\Gamma$, showing collective motion of Li2 atoms (orange). The blue and grey arrows indicate mode displacement directions.}
\end{figure}
 
To gain insight into the physical nature of ND and XRD refined sites in $\beta$-\ch{Li3PS4}, we examine two low-energy structures that were previously proposed as the ground state in separate first-principles studies \cite{lim_beta_lps_diffusion_abinit_2018, yang_elastic_2016}. These highly similar structures are shown in Figures \ref{fig:lps_b_hsgs_struc} and \ref{fig:lps_b_lsgs_struc}. One contains fully occupied Li1' (8d) and Li2' (4b) sites, which yields well-ordered linear chains of Li1' and Li2' atoms along [010] and [001], and retains the Pnma symmetry of the underlying lattice---we will refer to this as the XRD ground state (XRD-GS) (Figure \ref{fig:lps_b_hsgs_struc}). The other proposed structure is the true, lowest energy ground state in our data set, which is reported to have fully occupied Li1' (8d) and Li2' (4b) sites, but the square planar coordinated Li2' atoms are displaced off-center to a neighboring 5-fold coordination environment, characteristic of the ND refined Li2 (8d) site---we will refer to this as the ND ground state (ND-GS) (Figure \ref{fig:lps_b_lsgs_struc}). The Li2 chain of atoms in ND-GS is staggered along [010], which leads to decreased symmetry (P$2_12_12_1$) compared to XRD-GS (Pnma). The Li site fractional occupancies of ND-GS can be described in the basis of the ND refined sites as x$_{\text{Li1}} = 1$, x$_{\text{Li2}} = 0.5$, and x$_{\text{Li3}} = \text{x}_{\text{Li4}} = 0$.

Although Li2' (4b) is located merely 0.6 $\AA$ from a neighboring Li2' (8d) site, the decrease in site energy is substantial, as ND-GS is 3.4 meV/atom lower than XRD-GS. Furthermore, when comparing phonon dispersion spectra, we find that XRD-GS is dynamically unstable with 2 nearly degenerate optical imaginary modes (Figure \ref{fig:lps_b_hsgs_band}), while ND-GS is dynamically stable with no imaginary modes (Figure \ref{fig:lps_b_lsgs_band}), agreeing well with previous reports \cite{lim_beta_lps_diffusion_abinit_2018}. When visualizing the XRD-GS imaginary optical modes at the $\Gamma$ wave vector, we observe a collective motion of Li2' atoms (Figure \ref{fig:lps_b_hsgs_mode}). This indicates that the XRD refined Li2' (4b) site is a high energy transition state for Li hopping between two neighboring Li2 (8d) sites. These findings highlight the importance of distinguishing fine details of the Li sublattice, as substantial differences in physical behavior can arise when site locations are slightly perturbed.
\begin{figure}
    \centering
    \begin{subfigure}[b]{0.48\textwidth}
        \centering
        \includegraphics[scale=0.52]{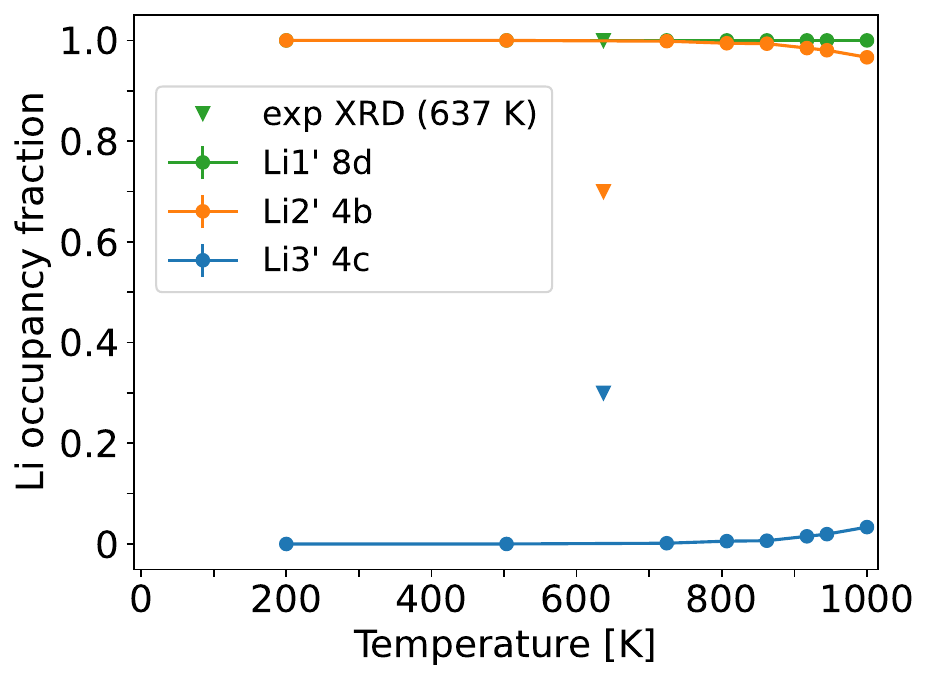}
        \caption{}
        \label{fig:li3ps4_b_xrd_occu}
    \end{subfigure}
    \hspace{0.1cm}
    \begin{subfigure}[b]{0.48\textwidth}
        \centering
        \includegraphics[scale=0.52]{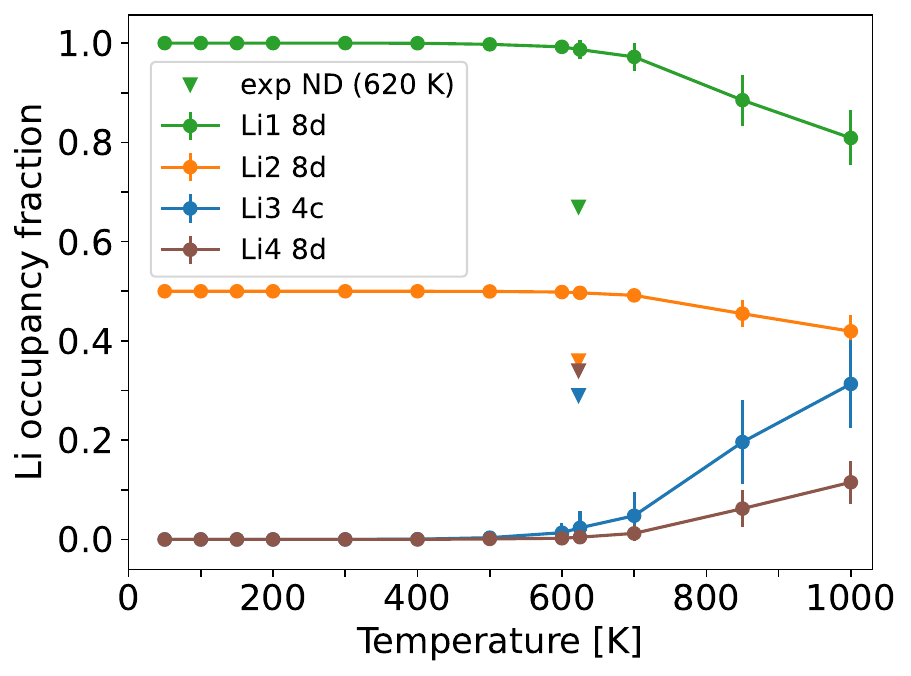}
        \caption{}
        \label{fig:li3ps4_b_npd_occu}
    \end{subfigure}
    
    \caption{Li site fractional occupancies in $\beta$-\ch{Li3PS4} across temperature in MC simulations, for the a) XRD-refined lattice and b) ND-refined lattice.}
    \label{fig:li3ps4_b_occus}
\end{figure}

The thermodynamic disordering behavior of the XRD and ND refined structures at elevated temperature are also compared. We fit separate cluster expansions on each lattice and perform MC simulations to predict the Li site disorder as a function of temperature. In Figure \ref{fig:li3ps4_b_occus}, the Li site fractional occupancies across temperature are plotted. The XRD structure begins to disorder from XRD-GS at approximately 900 K, and by 1000 K changes in the Li fractional occupancies are still relatively small, yielding poor agreement with the experimental XRD refinement (Figure \ref{fig:li3ps4_b_xrd_occu}). The ND structure begins to disorder from ND-GS at a lower temperature of about 600 K, and by 1000 K has significant changes in its Li fractional occupancies, highlighted by Li1 (8d) and Li3 (4c) having occupancies of 0.8 and 0.3, respectively. These values show reasonable agreement with the ND refinement at 620 K (0.7 and 0.3) (triangles in Figure \ref{fig:li3ps4_b_npd_occu}). Our simulations on both the XRD and ND structures underestimate the experimentally reported configurational disorder. However, the ND structure is predicted to have greater disorder and thus better agreement with experiment, suggesting that the ND refinement is more accurate. Specifically, introducing the Li4 (8d) site and increasing multiplicity of Li2' (4b) to Li2 (8d) generates more configurational states that appear essential towards accurately describing the thermodynamics of this phase.

\subsubsection{$\alpha$-\ch{Li3PS4} structure}

\begin{figure}[t!]
    \centering
    \begin{subfigure}[b]{0.48\linewidth}
        \centering
        \includegraphics[scale=0.45]{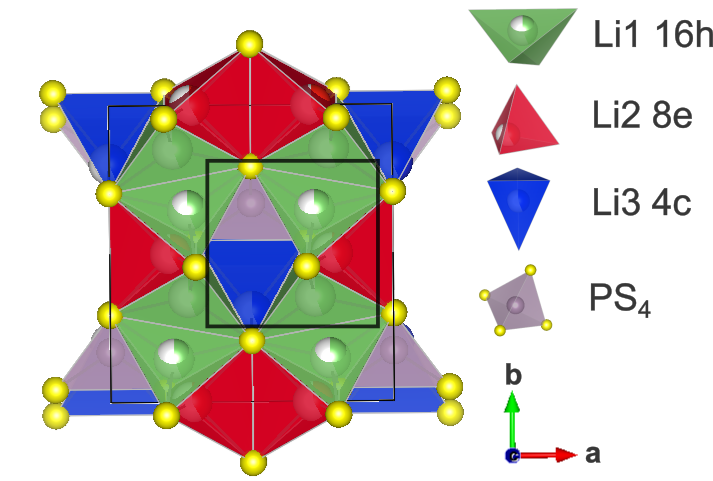}
        \caption{}
        \label{fig:a_li3ps4_unit}
    \end{subfigure}
    \hspace{0.1cm}
    \begin{subfigure}[b]{0.48\linewidth}
        \centering
        \includegraphics[scale=0.3]{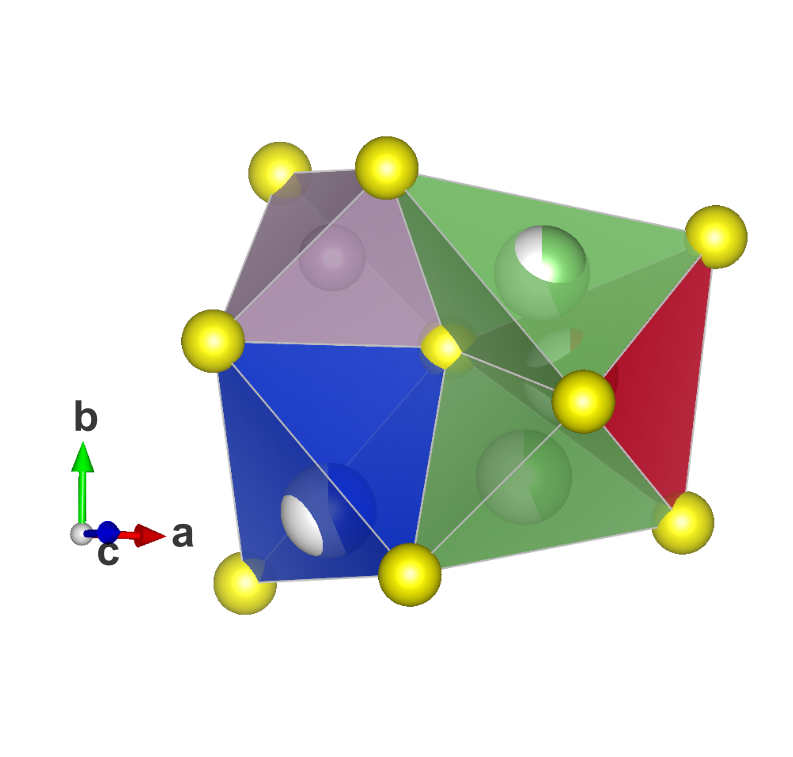}
        \caption{}
        \label{fig:a_li3ps4_local}
    \end{subfigure}
    \hspace{0.1cm}
    \begin{subfigure}[b]{0.44\linewidth}
        \centering
        \includegraphics[scale=0.33]{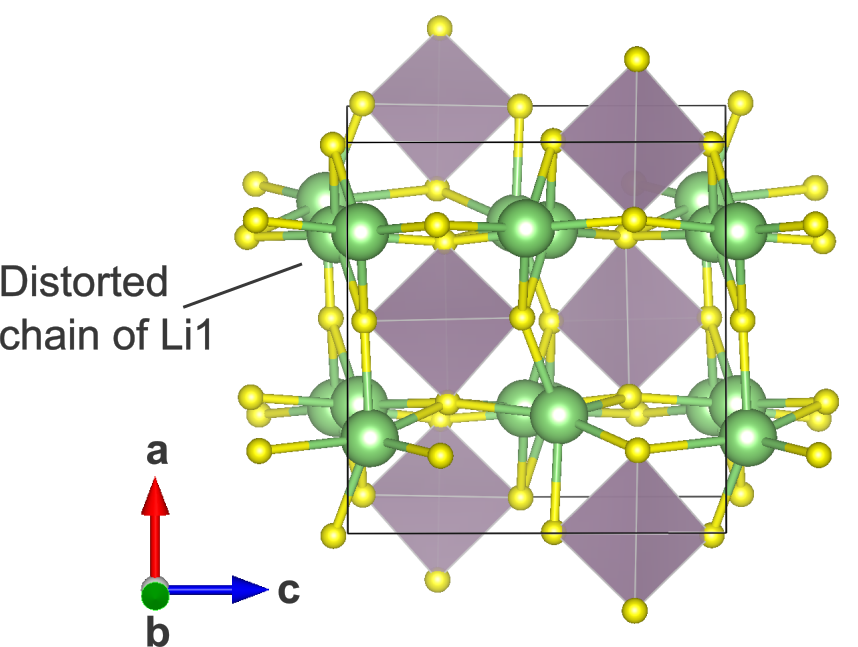}
        \caption{}
        \label{fig:a_li3ps4_gs}
    \end{subfigure}
    \hspace{0.1cm}
    \begin{subfigure}[b]{0.52\textwidth}
        \centering
        \includegraphics[scale=0.56]{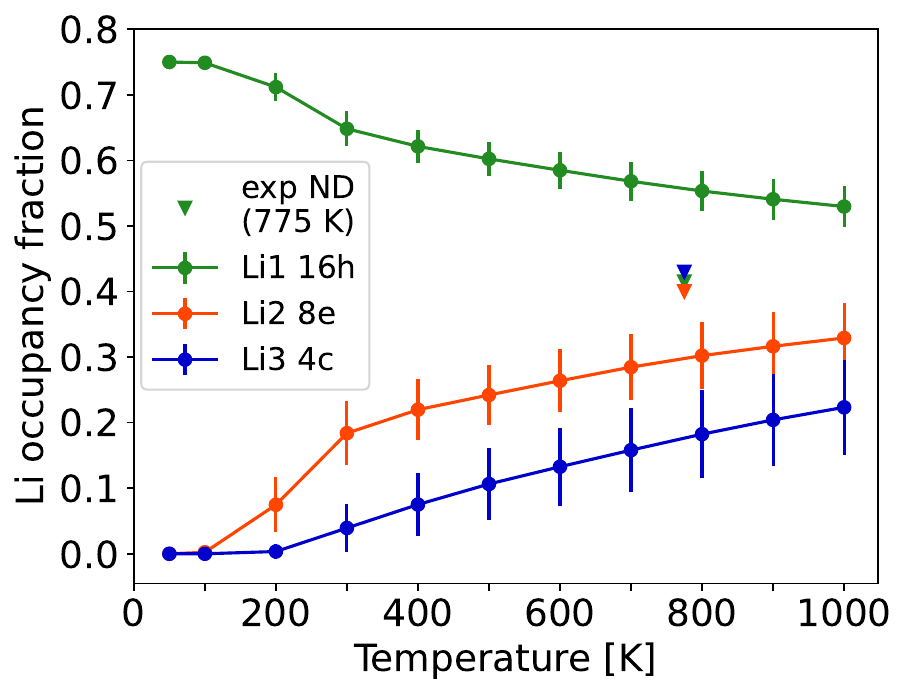}
        \caption{}
        \label{fig:li3ps4_a_npd_occu}
    \end{subfigure}
    \caption{Structure of $\alpha$-\ch{Li3PS4}. a) The unit cell with Li1 (16h) (green), Li2 (8e) (red), and Li3 (4c) (blue) sites. The boxed region is shown in greater detail in b) to display the local Li coordination. Nearest Li1 and Li2 sites face-share to form a connected Li channel along [010]. Two adjacent channels are connected by Li3 sites, which edge-share with Li1. c) The ground state structure with monoclinic P$2_1$/c symmetry. Li atoms occupy only Li1 sites, forming distorted linear chains of Li along [010] and [001]. d) Li site fractional occupancies from MC simulations.}
\end{figure}

At high temperature (T $> 725$ K), $\beta$ transforms to the orthorhombic Cmcm $\alpha$-\ch{Li3PS4}, increasing symmetry (Cmcm is a supergroup of Pnma) and slightly decreasing in density (1.6\%) \cite{kaup_li3ps4_nd_2020}. ND refinements report a Li sublattice containing Li1 (16h), Li2 (8e), and Li3 (4c) sites with high degree of disorder, as indicated by the isotropic Li fractional occupancies of around 0.4 \cite{kaup_li3ps4_nd_2020}. An earlier refinement with XRD was deemed inconclusive, as only 1/3 Li atoms in the formula unit were refined to 1 distinct site, and there were large errors in the atomic displacement parameter (ADP) \cite{kanno_alpha_li3ps4_structure}. The ND refinement shows significant improvement by locating 2.9/3 Li and containing lower error in ADP \cite{kaup_li3ps4_nd_2020}. Therefore, we use the ND-refined structure, which contains 3 tetrahedral Li sites over which Li atoms can disorder, to construct our cluster expansion for $\alpha$-\ch{Li3PS4}. The disordered unit cell and local Li coordination of $\alpha$-\ch{Li3PS4} are shown in Figures \ref{fig:a_li3ps4_unit} and \ref{fig:a_li3ps4_local}, respectively.
We can observe that $\alpha$-\ch{Li3PS4} contains a well-connected 1D channel of face-sharing Li1-Li2-Li1 sites along [010] (Figure \ref{fig:a_li3ps4_local}), which can be associated with fast Li-ion conduction \cite{wang_design_2015}. The Li3 sites, which edge-share with Li1, serve to bridge adjacent Li1-Li2-Li1 channels.

From MC simulated annealing, we find the ground state of $\alpha$-\ch{Li3PS4} to be 3.2 meV/atom (26 meV/f.u.) above the ground state of the $\beta$ polymorph, and 8.0 meV/atom (64 meV/f.u.) above the $\gamma$ polymorph. The ground state of $\alpha$-\ch{Li3PS4} is shown in Figure \ref{fig:a_li3ps4_gs}, which contains a slight monoclinic distortion (lattice angle $\gamma$ = 86.4$\degree$). Li atoms only occupy the Li1 (16h) sites and form distorted linear Li chains along [010] and [001] (Figure \ref{fig:a_li3ps4_gs}). This indicates that Li1 (16h) sites are the most stable, and their face-sharing Li2 (8e) neighbors are higher energy intermediate sites that facilitate rapid Li diffusion. Similarly, the Li3 (4c) sites are higher energy intermediate sites that connect adjacent Li1-Li2-Li1 channels and promote 3D conductivity \cite{kaup_li3ps4_nd_2020}.

MC simulations show that Li starts to occupy Li2 (8e) sites at 200 K, and Li3 (4c) sites at 300 K (Figure \ref{fig:li3ps4_a_npd_occu}). $\alpha$-\ch{Li3PS4} thus begins to disorder at a much lower temperature compared to $\beta$-\ch{Li3PS4}. By 600 K, Li atoms already occupy a significant fraction of each Li site, whereas $\beta$-\ch{Li3PS4} only begins to disorder at this temperature. Thus, the $\alpha$ polymorph contains much greater configurational disorder compared to $\beta$. This is in qualitative agreement with the experimental ND refinement, which shows very isotropic Li fractional occupancies of around 0.4 for each site at 775 K (triangles in Figure \ref{fig:li3ps4_a_npd_occu}).

\subsection{\ch{Li3PS4} phase stability}

\begin{figure}
    \centering
    \includegraphics[scale=0.7]{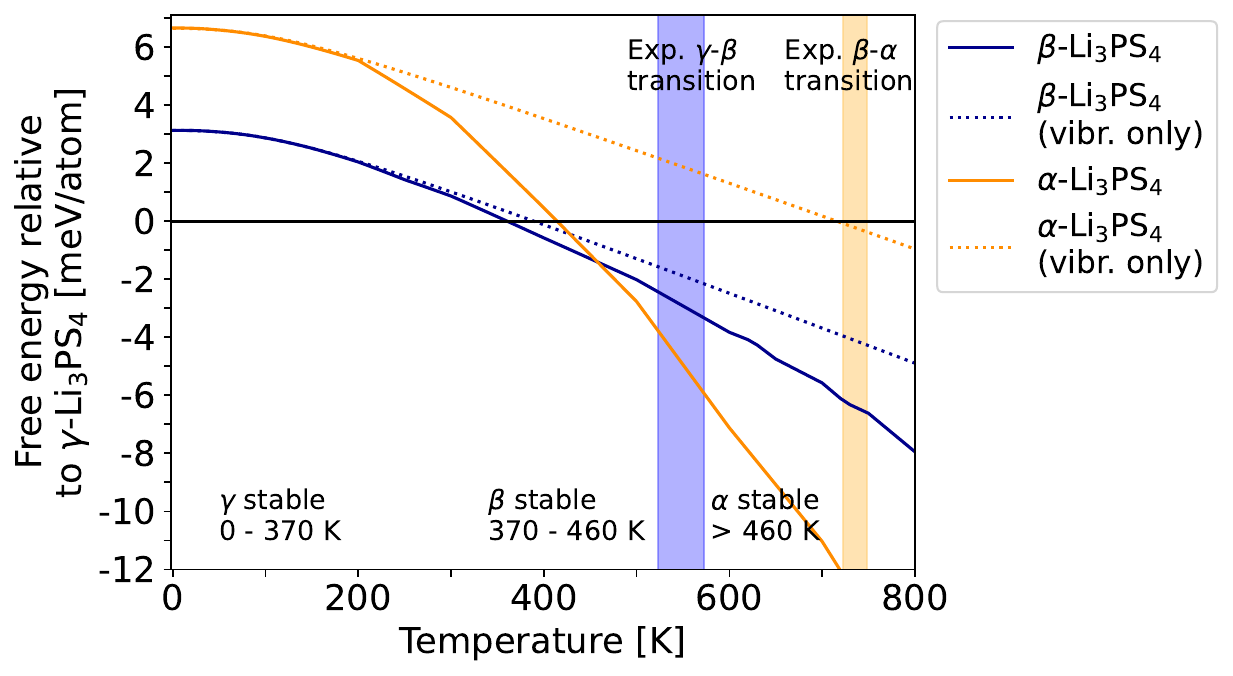}
    \caption{Free energy of $\alpha$ and $\beta$-\ch{Li3PS4} relative to $\gamma$-\ch{Li3PS4}. All experimentally reported phase transitions are observed when accounting for all free energy contributions (solid lines). $\gamma$-$\beta$ transition occurs at 370 K (523-573 K from exp.), $\beta$-$\alpha$ at 460 K (723-748 K from exp), and $\alpha$-$\gamma$ at 420 K (533 K from exp upon cooling). Only the $\gamma$-$\beta$ transition is observed when including only vibrational entropy contributions (dotted lines). Shaded blue and orange regions indicate experimentally observed $\gamma$-$\beta$ and $\beta$-$\alpha$ phase transitions, respectively.}
    \label{fig:li3ps4_free_ens}
\end{figure}

Using the structural models we validated for the \ch{Li3PS4} polymorphs, we assess the stability of each polymorph across temperature by calculating and comparing their free energy. In Figure \ref{fig:li3ps4_free_ens}, we plot the free energy of $\alpha$ and $\beta$ relative to $\gamma$-\ch{Li3PS4}. Since $\gamma$ contains well-ordered Li, we assume it to only create vibrational entropy. At 0 K, the polymorphs ranked in order of decreasing stability are $\gamma$, $\beta$, and $\alpha$. The $\gamma$-$\beta$ transition is predicted to occur at 370 K and the $\beta$-$\alpha$ transition occurs at 460 K (Figure \ref{fig:li3ps4_free_ens}). This order of phase transitions matches with experiments, though the predicted transition temperatures are 200-300 K below experimentally observed values. In experiments, it is also commonly observed that $\alpha$ directly transforms to $\gamma$ without forming $\beta$ upon cooling \cite{homma_li3ps4_crystal_2011, kaup_li3ps4_nd_2020}, which we predict would occur at 420 K. At this temperature, the free energy differences among the polymorphs are very small ($< 1$ meV/atom), which helps rationalize why a direct transition can occur, especially if the transformation to $\gamma$ is more kinetically favorable than forming $\beta$. We note that the r$^2$SCAN density functional \cite{furness_r2scan_2020} is required to predict the correct order of \ch{Li3PS4} polymorph stability, since $\gamma$ is predicted to be unstable across all temperatures when using PBE \cite{perdew_pbe_1996} (SI Figure S1), which was also reported in previous first-principles calculations \cite{zhou_entropically_2019}.

When configurational entropy contributions are neglected (dotted lines in Figure \ref{fig:li3ps4_free_ens}), the free energy of $\alpha$ always lies above $\beta$, such that the only accessible transition is $\gamma$-$\beta$. This is attributed to the highly similar vibrational free energy profiles of $\alpha$ and $\beta$. After $\alpha$ begins to disorder at around 200 K, its configurational entropy increases faster than $\beta$, which drives the increased stability of $\alpha$ at high temperature. Furthermore, the exclusion of configurational entropy only slightly increases the $\gamma$-$\beta$ transition temperature to 390 K, since $\beta$ has low configurational entropy at this temperature. The main source of stability for $\beta$-\ch{Li3PS4} is thus vibrational entropy.

\begin{figure}[t!]
\centering
    \begin{subfigure}[b]{0.47\linewidth}
        \centering
        \includegraphics[scale=0.53]{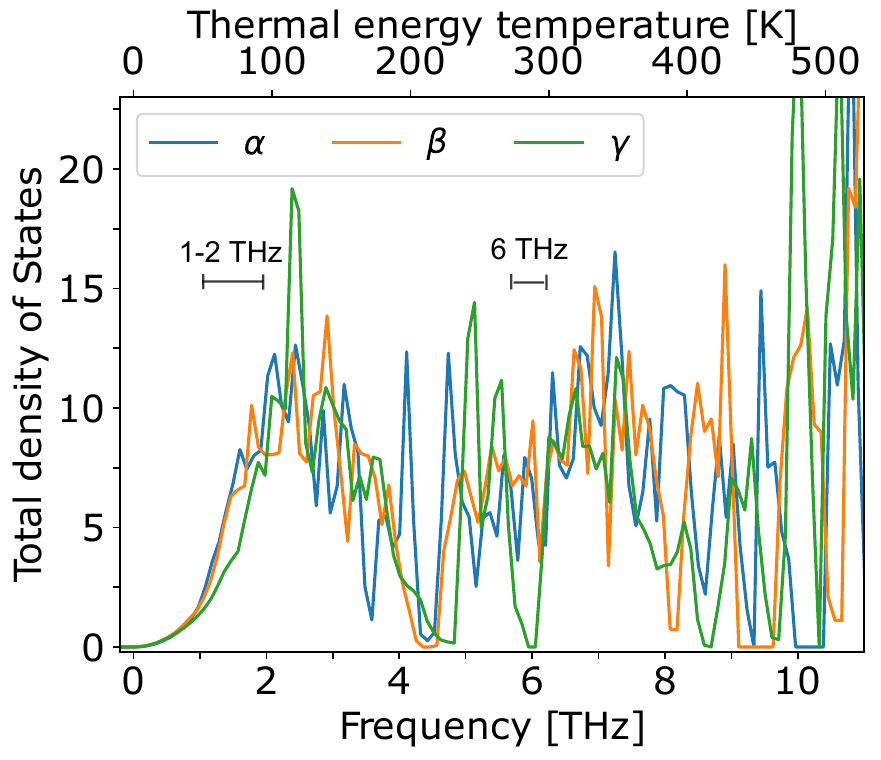}
        \caption{Total DOS}
        \label{fig:li3ps4_phon_tdos}
    \end{subfigure}
    \hspace{0.2cm}
    \begin{subfigure}[b]{0.47\linewidth}
        \centering
        \includegraphics[scale=0.53]{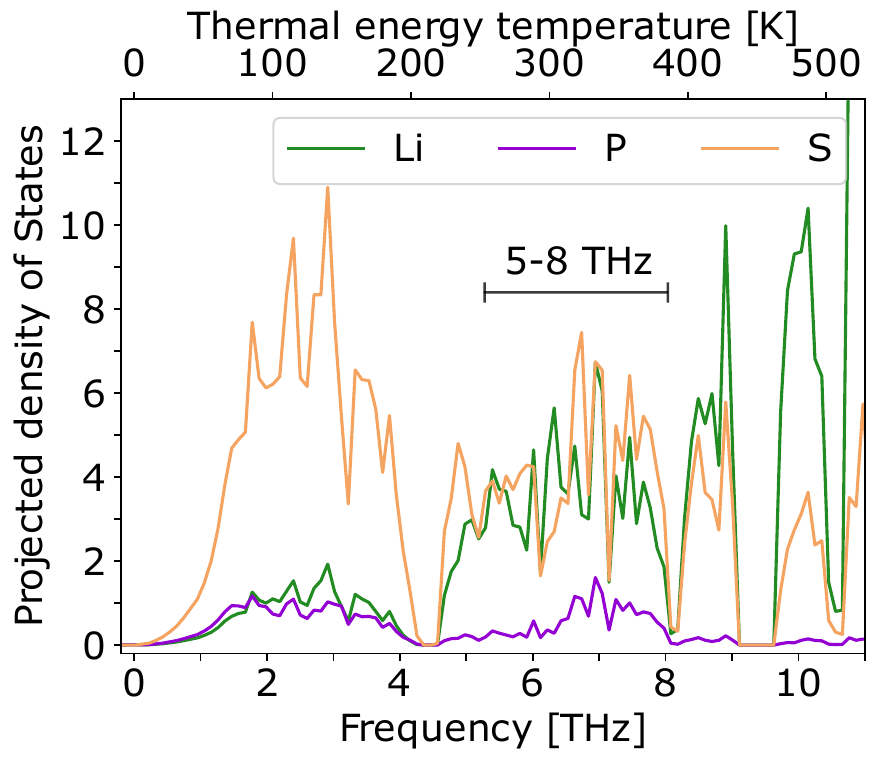}
        \caption{Projected DOS $\beta$-\ch{Li3PS4}}
        \label{fig:li3ps4_b_phon_pdos}
    \end{subfigure}
    \hspace{0.2cm}
    \begin{subfigure}[b]{0.47\linewidth}
        \centering
        \includegraphics[scale=0.53]{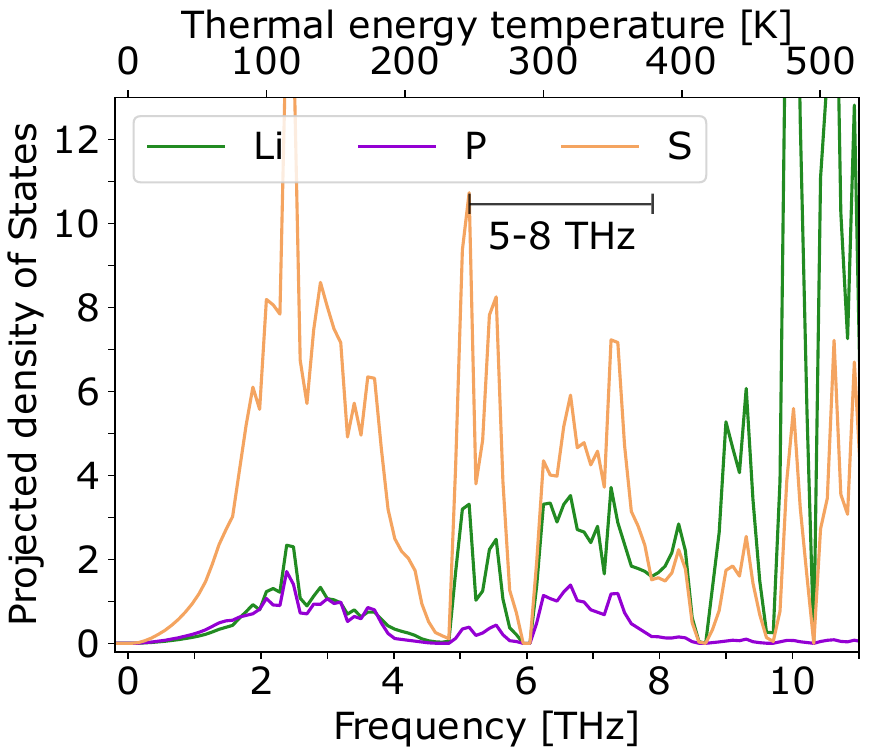}
        \caption{Projected DOS $\gamma$-\ch{Li3PS4}}
        \label{fig:li3ps4_g_phon_pdos}
    \end{subfigure}
    \hspace{0.2cm}
    \begin{subfigure}[b]{0.47\linewidth}
        \centering
        \includegraphics[scale=0.53]{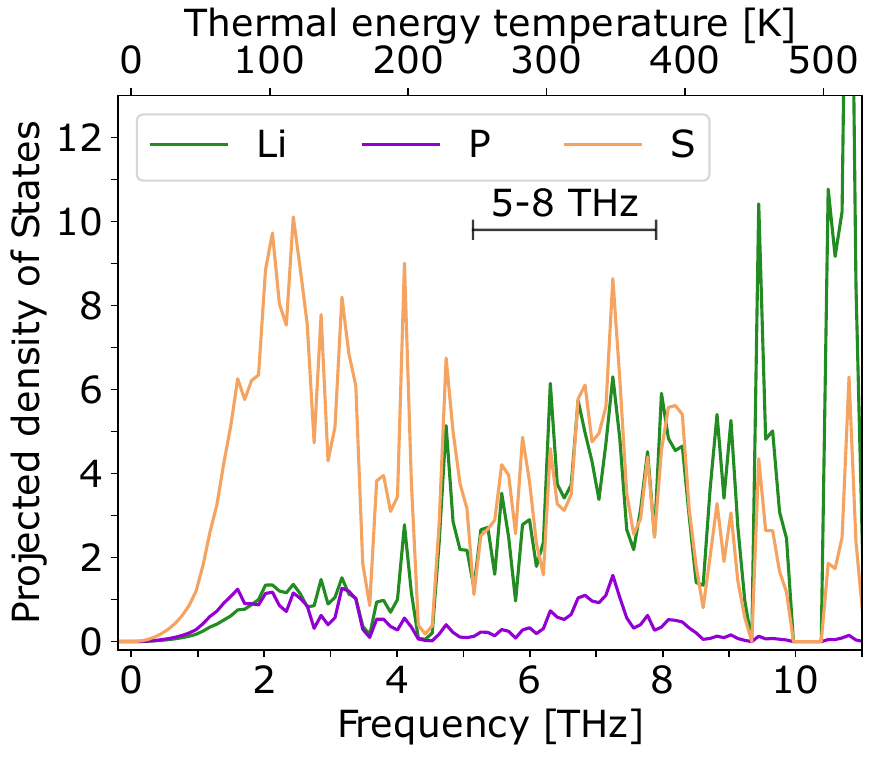}
        \caption{Projected DOS $\alpha$-\ch{Li3PS4}}
        \label{fig:li3ps4_a_phon_pdos}
    \end{subfigure}
\caption{Phonon density of states. a) Total density of states (TDOS) for each \ch{Li3PS4} polymorph. Projected density of states (PDOS) onto Li, P, and S atoms for b) $\beta$, c) $\gamma$ and d) $\alpha$-\ch{Li3PS4} as a function of frequency and thermal energy temperature ($T = hf/{k_B}$), normalized per unit cell of $\beta$-\ch{Li3PS4} (4 formula units). $\gamma$-\ch{Li3PS4} contains smaller DOS at frequencies of 1-2 THz and 6 THz. The region of 1-2 THz is dominated by S modes. At 5-8 THz, $\gamma$ has much larger density of S modes compared to Li, while in $\alpha$ and $\beta$ there are equal contributions of each.}
\end{figure}

To rationalize the distinctly greater vibrational entropy in $\beta$ and $\alpha$ compared to $\gamma$-\ch{Li3PS4}, we compare the phonon density of states (DOS) in each phase, which are shown in Figure \ref{fig:li3ps4_phon_tdos}. $\beta$ and $\alpha$-\ch{Li3PS4} contain significantly larger DOS at 1-2 THz and around 6 THz (Figure \ref{fig:li3ps4_phon_tdos}). The projected DOS (pDOS) shows that for all phases, the 1-2 THz region is dominated by sulfur (S) modes, which are activated at low temperature around 100 K. From visualizing these modes, we observe that they mainly correspond to librations of the \ch{PS4} groups. Furthermore, $\gamma$ has no vibrations at 6 THz, whereas the high temperature phases contain significant DOS near this frequency. This frequency lies in the region between 5 to 8 THz (240 to 380 K), where $\beta$ (Figure \ref{fig:li3ps4_b_phon_pdos}) and $\alpha$ (Figure \ref{fig:li3ps4_a_phon_pdos}) have roughly equal projected density of Li and S phonon modes, whereas in $\gamma$ there is a significantly larger projected density of S modes than Li (Figure \ref{fig:li3ps4_g_phon_pdos}). The activation of larger amplitude Li modes at around room temperature contributes to greater thermodynamic stability, and potentially towards high Li mobility in $\beta$ and $\alpha$-\ch{Li3PS4}. This finding is consistent with prior reports highlighting the relation between fast Li-ion conductivity and vibrational entropy in some superionic conductors \cite{muy_sic_phonon}.
 
\begin{figure}
    \centering
    \includegraphics[scale=0.8]{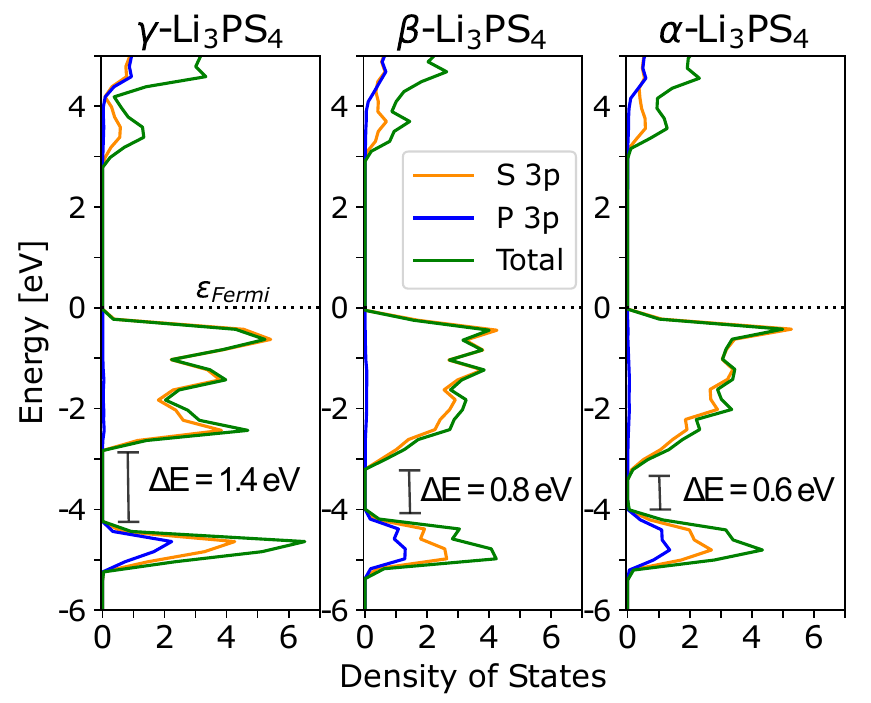}

    \caption{Orbital resolved electronic density of states of $\gamma$, $\beta$, and $\alpha$-\ch{Li3PS4} ground state structures. All phases exhibit a valence band manifold consisting of mainly S 3p states, and a lower energy core band manifold with S 3p and P 3p states. $\gamma$-\ch{Li3PS4} has a larger energy gap ($\Delta$E) between valence and core band manifolds (1.4 eV) compared to $\alpha$ and $\beta$ (0.6 and 0.8 eV). This leads to a smaller energy spread in the valence and core band manifolds in $\gamma$. }
    \label{fig:li3ps4_e_dos}
\end{figure}

Since the differences in low-temperature vibrational modes are likely dictated by the bonding within the S sublattice, we examine the electronic density of states of each ground state, which are shown in Figure \ref{fig:li3ps4_e_dos}. For all three polymorphs, the manifold of valence bands below the Fermi level dominantly consists of S 3p states which are spread over an energy range of $\sim$3 eV. The relatively large band widths indicate that these states are delocalized in character and should represent long-range van der Waals interactions among S atoms in separate \ch{PS4} units (Figure \ref{fig:li3ps4_e_dos}). The lower energy core band manifold consists of mostly P 3p and S 3p states, which we attribute to P-S binding between the \ch{PS4} groups. The core band states are spread over a narrower energy range of $\sim$1 eV, indicating that these states are more localized.

A key difference in electronic structure is observed in the $\gamma$ polymorph, which has a larger energy gap between the core and valence band states, arising from narrower band widths in the core and valence band manifolds (Figure \ref{fig:li3ps4_e_dos}). The narrower core band widths can arise from stronger hybridization of S 3p and P 3p states in neighboring \ch{PS4} units, leading to more localization. This stronger hybridization between S and P atoms may lead to smaller interaction between S 3p states on neighboring \ch{PS4} groups, which contributes to decreased valence band widths. It appears that the uni-directional \ch{PS4} arrangement and denser hcp-type anion packing in $\gamma$-\ch{Li3PS4}\cite{kaup_li3ps4_nd_2020} facilitates more isotropic and localized P-S bonding states to inhibit facile S motion. These factors would contribute to greater S sublattice stiffness and reduced density of low-frequency S vibrational modes.

\subsection{\ch{Li7PS6} polymorphs}

\begin{figure}[t!]
    \centering
    \begin{subfigure}[b]{0.45\linewidth}
        \centering
        \includegraphics[scale=0.3]{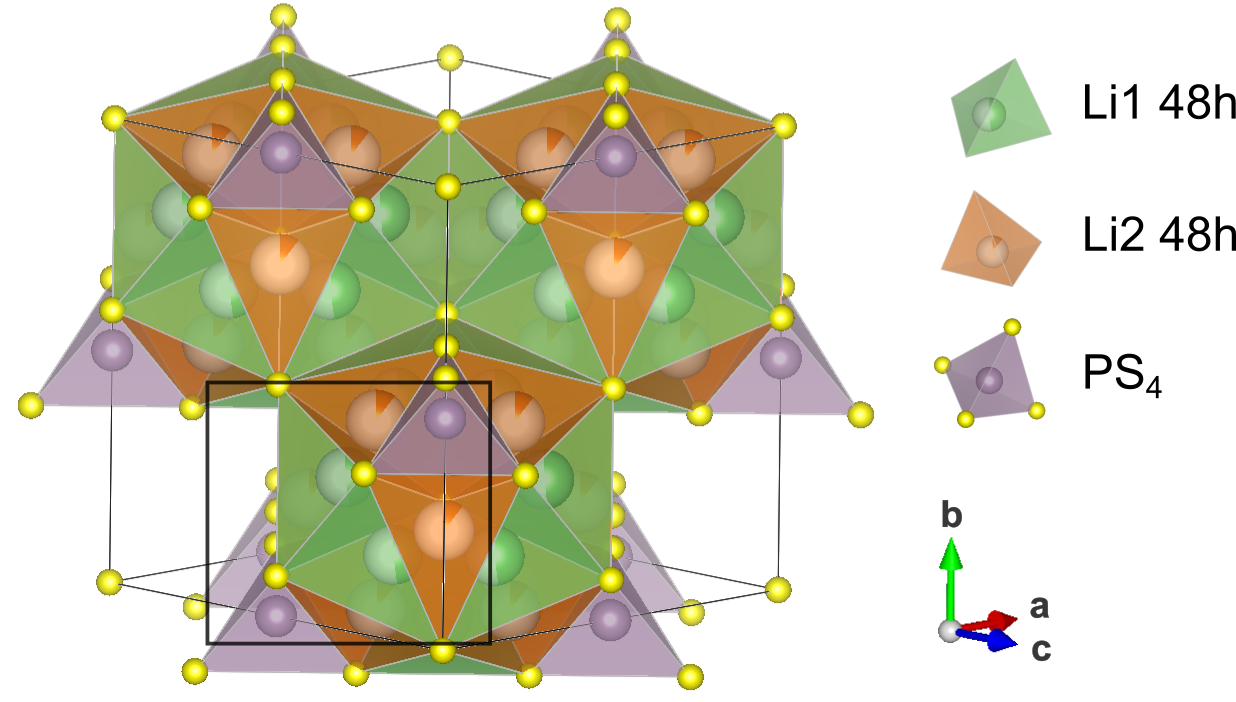}
        \caption{Unit cell}
        \label{fig:ht_li7ps6_unit}
    \end{subfigure}
    \hspace{1.3cm}
    \begin{subfigure}[b]{0.45\linewidth}
        \centering
        \includegraphics[scale=0.29]{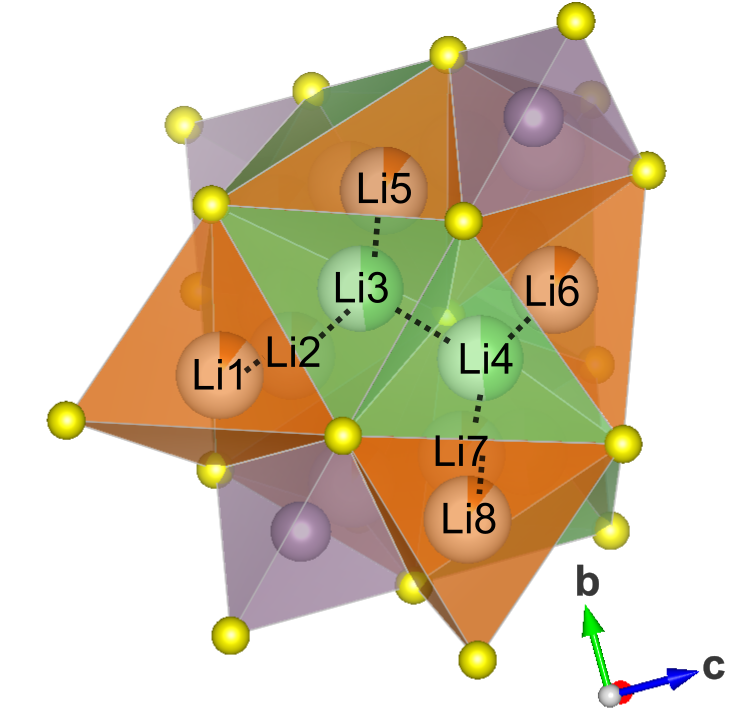}
        \caption{Local Li substructure}
        \label{fig:ht_li7ps6_local}
    \end{subfigure}
    
    \begin{subfigure}[b]{0.45\linewidth}
        
        \includegraphics[scale=0.55]{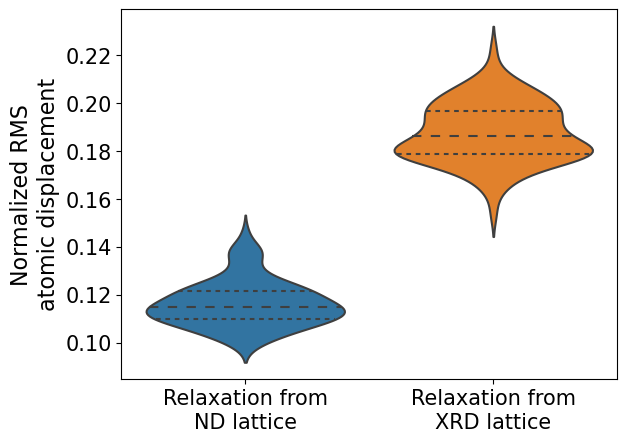}
        \caption{Distributions of DFT relaxations from ND and XRD refinements}
        \label{fig:ht_li7ps6_nd_xrd}
    \end{subfigure}
    \hspace{1cm}
    \begin{subfigure}[b]{0.45\textwidth}
        \centering
        \includegraphics[scale=0.27]{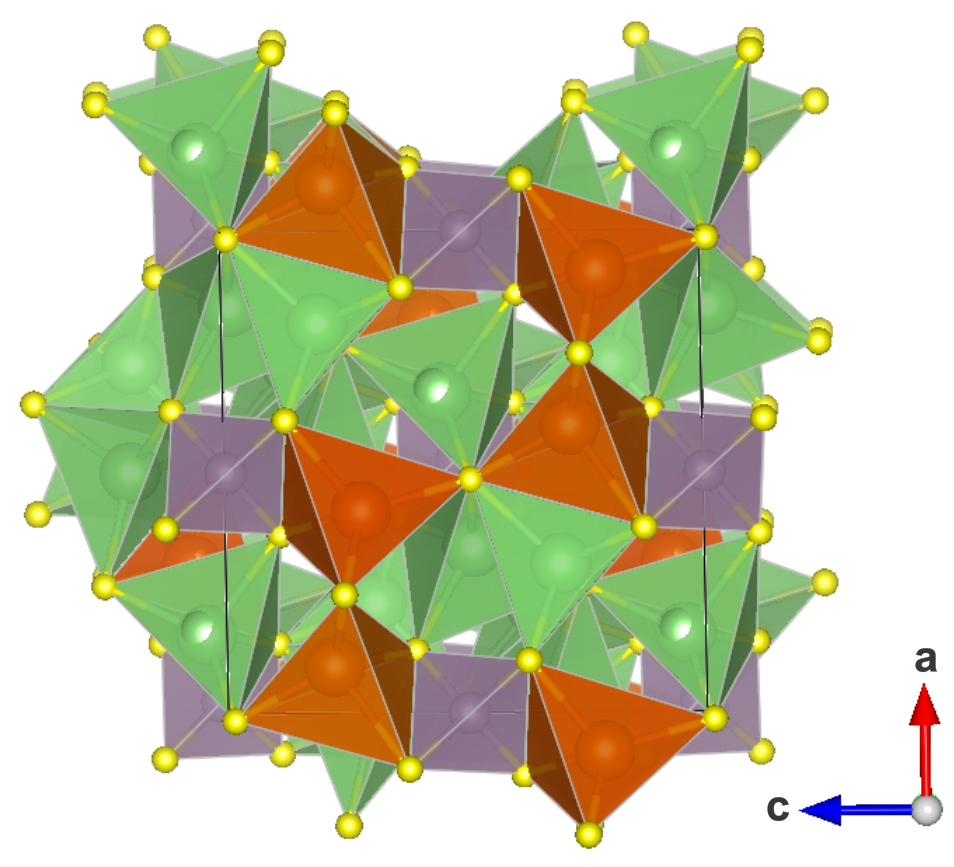}
        \caption{Ground state structure}
        \label{fig:ht_li7ps6_gs}
    \end{subfigure}
    
    \caption{Structure of HT-\ch{Li7PS6}. a) The unit cell (F-43m) with Li1 (48h) (green) sites that corner-share with \ch{PS4}, and Li2 (48h) (orange) sites that edge-share with \ch{PS4}. The boxed region containing a single cage-like Li substructure is shown in greater detail in b), which displays the local Li site coordination. Dotted lines connect face-sharing Li atoms. c) Violin plots displaying distributions of normalized root mean squared (NRMS) atomic displacements of DFT relaxed configurations from ND and XRD refinements. Dashed lines denote division between data quartiles. d) Ground state structure of HT-\ch{Li7PS6}, containing 6 occupied Li2 (48h) sites (orange).}
\end{figure}

Experiments show that orthorhombic LT-\ch{Li7PS6} (Pna$2_1$) is well-ordered and transforms to the higher symmetry cubic HT-\ch{Li7PS6} (F-43m) phase at 483 K \cite{kong_2020_li7ps6_xrd}. According to XRD refinements, HT-\ch{Li7PS6} contains a disordered Li sublattice with one distinct Li1 (48h) site, which corner-shares with \ch{PS4} units and face-shares with its nearest Li1 neighbor \cite{kong_2020_li7ps6_xrd}. No ND refinement has yet been reported on the HT-\ch{Li7PS6} phase; however, ND refinements have been reported for a Cl-doped analogue \ch{Li6PS5Cl} \cite{schlenker_li6ps5cl_nd_pdf}. An additional Li2 (48h) site was identified in \ch{Li6PS5Cl} that edge-shares with \ch{PS4} units, and face-shares with its nearest Li1 and Li2 neighbors to form a cage-like Li substructure (Figure \ref{fig:ht_li7ps6_local}), while the sublattices of the \ch{PS4} and isolated \ch{S} atoms are identical. Since Cl substitutes a fraction of S atoms without causing much change in lattice parameters, we presume that the Li sites in the doped and pristine phases are very similar and comparable. \cite{schlenker_li6ps5cl_nd_pdf}.

As done with the $\beta$-\ch{Li3PS4} phase, we compare the NRMS atomic relaxations (Equation \ref{nrms}) of the DFT relaxed configurations starting from either the ND or XRD refined atomic positions of HT-\ch{Li7PS6}, the distributions of which are plotted in Figure \ref{fig:ht_li7ps6_nd_xrd}. We observe that there is a much smaller NRMS atomic displacement from the ND-refined lattice (mean of 0.12) compared to the XRD-refined lattice (mean of 0.19), indicating that the ND positions for Li are closer to the energy minimum. 

We model Li-vacancy disorder in HT-\ch{Li7PS6} by fitting a CE using the ND refinement of \ch{Li6PS5Cl} containing Li1 (48h) and Li2 (48h) sites, with all Cl atoms replaced by S atoms. Through MC simulated annealing, we identify a ground state ordering, shown in Figure \ref{fig:ht_li7ps6_gs}, which contains 6 Li atoms in the unit cell (out of 28 Li) occupying Li2 sites, as evidenced by their edge-sharing with \ch{PS4} (orange in Figure \ref{fig:ht_li7ps6_gs}). The prominence of Li2 as a stable site in the ground state provides further evidence that the structure refined by ND is more accurate and that the Cl doping does not influence the location of Li sites.

\begin{figure}
    \centering
    \includegraphics[scale=0.6]{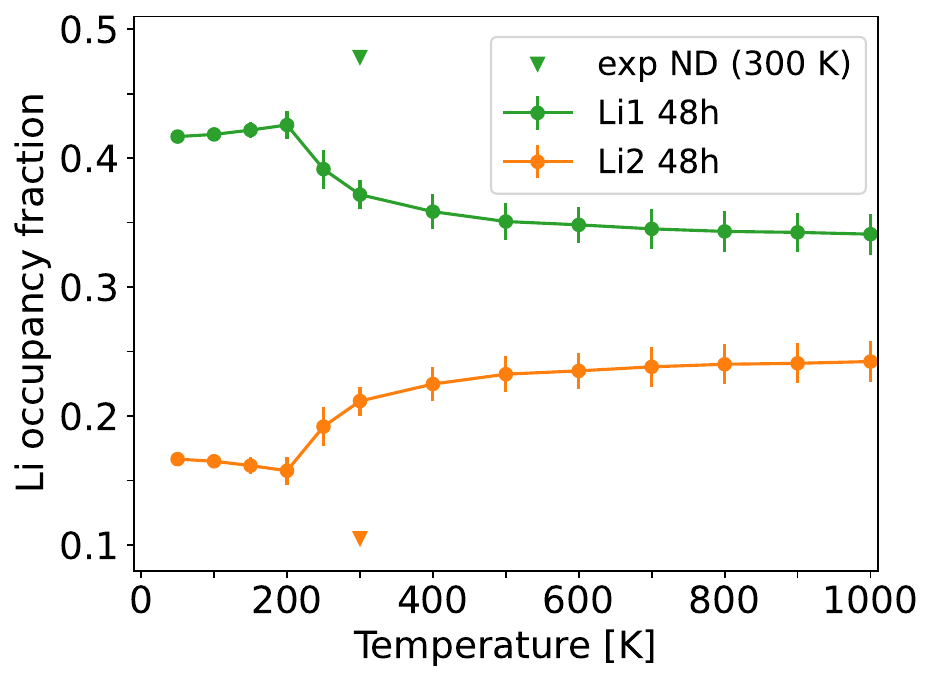}
    
    \caption{HT-\ch{Li7PS6} Li site occupancy fractions across temperature. The reported experimental occupancies (triangles)  were measured with ND on \ch{Li6PS5Cl} at 300 K, and are scaled by $7/6$ to account for the different Li stoichiometry, compared to \ch{Li7PS6}.}
    \label{fig:ht_li7ps6_occus}
\end{figure}

We perform MC simulations to predict the Li site occupancies as a function of temperature, which are plotted in Figure \ref{fig:ht_li7ps6_occus}. The fraction of Li occupying Li1 is greater at all simulated temperatures, in reasonable agreement with Li site occupancy of \ch{Li6PS5Cl} measured by ND at ambient temperature \cite{schlenker_li6ps5cl_nd_pdf}. The preference of Li going to Li1 sites could be explained by its corner-sharing with \ch{PS4}, which can reduce the repulsive interaction with P cations compared to the edge-sharing Li2 sites \cite{jun_cs_frame_2022}.

\begin{figure}
    \centering
    \begin{subfigure}[b]{0.45\linewidth}
        \centering
        \includegraphics[scale=0.5]{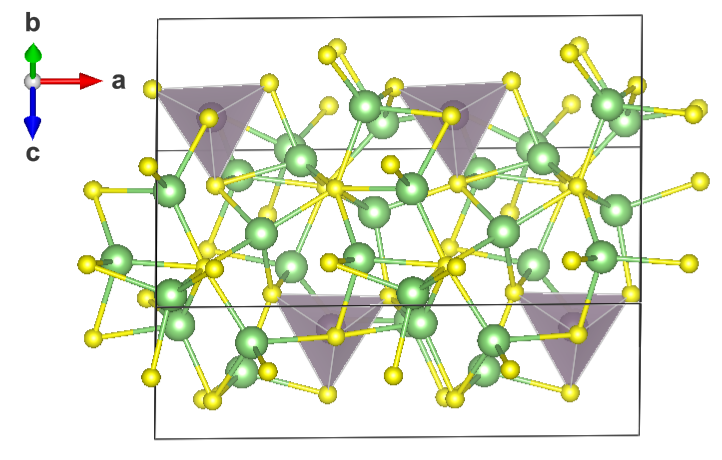}
        \caption{}
        \label{fig:lt_li7ps6_aimd_gs}
    \end{subfigure}
    \hspace{0.5cm}
    \begin{subfigure}[b]{0.45\linewidth}
        \includegraphics[scale=0.47]{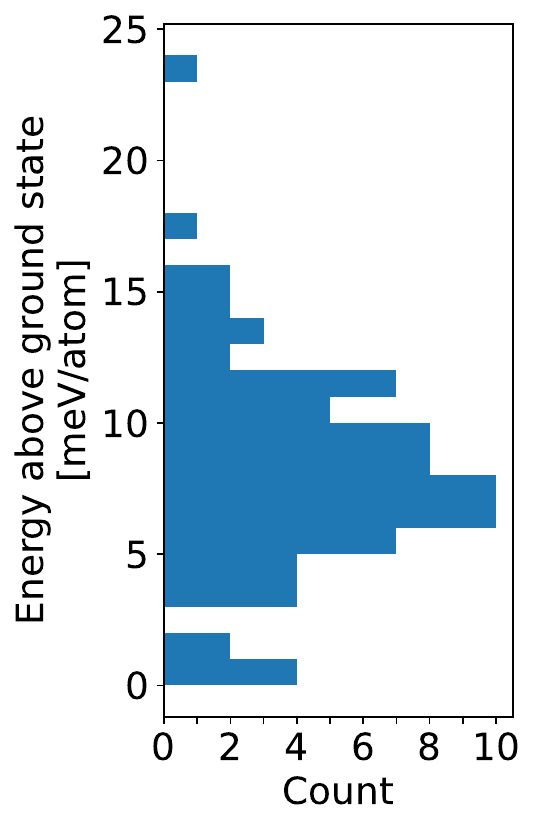}
        \centering
        \caption{}
        \label{fig:lt_li7ps6_aimd_ens}
    \end{subfigure}
    \caption{Structure of LT-\ch{Li7PS6}. a) The ground state structure identified from AIMD simulations. \ch{PS4} units relax slightly away from a parallel arrangement, contributing to a slight monoclinic distortion, with unit cell angles: ($\alpha$, $\beta$, $\gamma$) = (90.0, 91.0, 90.0). b) Histogram of energies above ground state for structures sampled along the AIMD trajectory. Reported energies are calculated from ionic relaxations using the PBE functional.}
    
\end{figure}

The HT-\ch{Li7PS6} ground state structure was found to be 10.4 meV/atom more stable than the ordered LT-\ch{Li7PS6} structure proposed by XRD (shown in SI Figure S2), suggesting that the XRD refinement for LT-\ch{Li7PS6} may not be accurate \cite{kong_2020_li7ps6_xrd}. To seek a more representative LT-\ch{Li7PS6} structure, we perturb its XRD refined structure through an AIMD simulation. The structure is heated to 800 K for 2 ps, held for 30 ps, and annealed to 100 K for 20 ps. Samples along the AIMD trajectory are relaxed, from which we identify a significantly more stable structure that is 1.2 meV/atom below the HT-\ch{Li7PS6} ground state. This new LT-\ch{Li7PS6} ground state (shown in Figure \ref{fig:lt_li7ps6_aimd_gs}) has a slight monoclinic distortion (lattice angle $\beta$ = 91\degree), resulting from a small relaxation of the \ch{PS4} units away from a parallel arrangement, and some Li are shifted to new coordination environments. We also find a large spread of energies among the sampled structures that were relaxed (Figure \ref{fig:lt_li7ps6_aimd_ens}), indicating that LT-\ch{Li7PS6} is likely configurationally disordered as well. We leave further analysis of the LT-\ch{Li7PS6} Li sublattice for future investigation. But our investigation confirms that, with our reassignment of the Li sites, LT-\ch{Li7PS6} is the ground state at low temperature.

\begin{figure}
    \centering
    \includegraphics[scale=0.6]{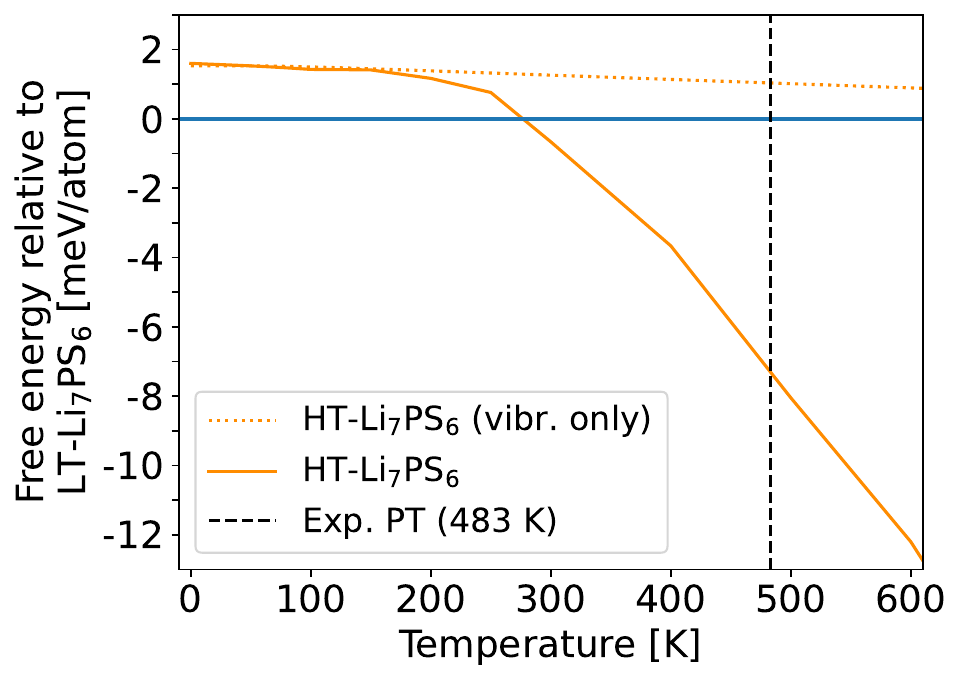}
    \caption{Free energy of HT-\ch{Li7PS6} relative to LT-\ch{Li7PS6}. Phase transition from LT to HT is predicted at 280 K, compared to 483 K from experiment \cite{kong_2020_li7ps6_xrd}.}
    \label{fig:ht_li7ps6_f_diff}
\end{figure}

Using our newly proposed ground states, we predict the phase stability of the \ch{Li7PS6} polymorphs at finite temperatures. In Figure \ref{fig:ht_li7ps6_f_diff}, we plot the free energy of HT-\ch{Li7PS6} relative to LT. HT-\ch{Li7PS6} becomes stable at 270 K, with the majority of its stabilization relative to LT-\ch{Li7PS6} arising from configurational entropy contributions (Figure \ref{fig:ht_li7ps6_f_diff}). Our predicted transition temperature is roughly 200 K below its experimentally observed value of 480 K. The likely cause for this understabilization of LT-\ch{Li7PS6} is that we may not have identified its true ground state yet, and that it contains significant configurational entropy contributions that have been neglected from our model because of the lack of a precise Li sublattice.
\subsection{\ch{Li7P3S11}}
\ch{Li7P3S11} crystallizes in the low symmetry P-1 space group, and is composed of \ch{PS4} and \ch{P2S7} units. In both XRD and ND refinements, the Li sublattice is ordered with 7 distinct sites \cite{onodera_li7p3s11_nd_xrd, yamane_li7p3s11_xrd_2007}. However, each refinement reports Li atoms occupying a different set of sites (the structures are shown in SI Figure S4). From our DFT calculations, we find that the XRD refined structure is substantially more stable than the ND refined structure by 9 meV/atom. A more recent first-principles study by Chang and coworkers proposed a disordered Li sublattice with 8 additional Li sites, identified from AIMD simulations \cite{chang_7311_aimd}. The authors enumerated structures based on the disordered Li sublattice and reported a ground state (SI Figure S4c) that is 16 meV/f.u. more stable than the XRD-refined structure, using the PBE functional. This value is qualitatively consistent with our calculations using r$^2$SCAN, which yield an energy difference of 22 meV/f.u. (1 meV/atom).
\begin{figure}
    \centering
     \includegraphics[scale=0.35]{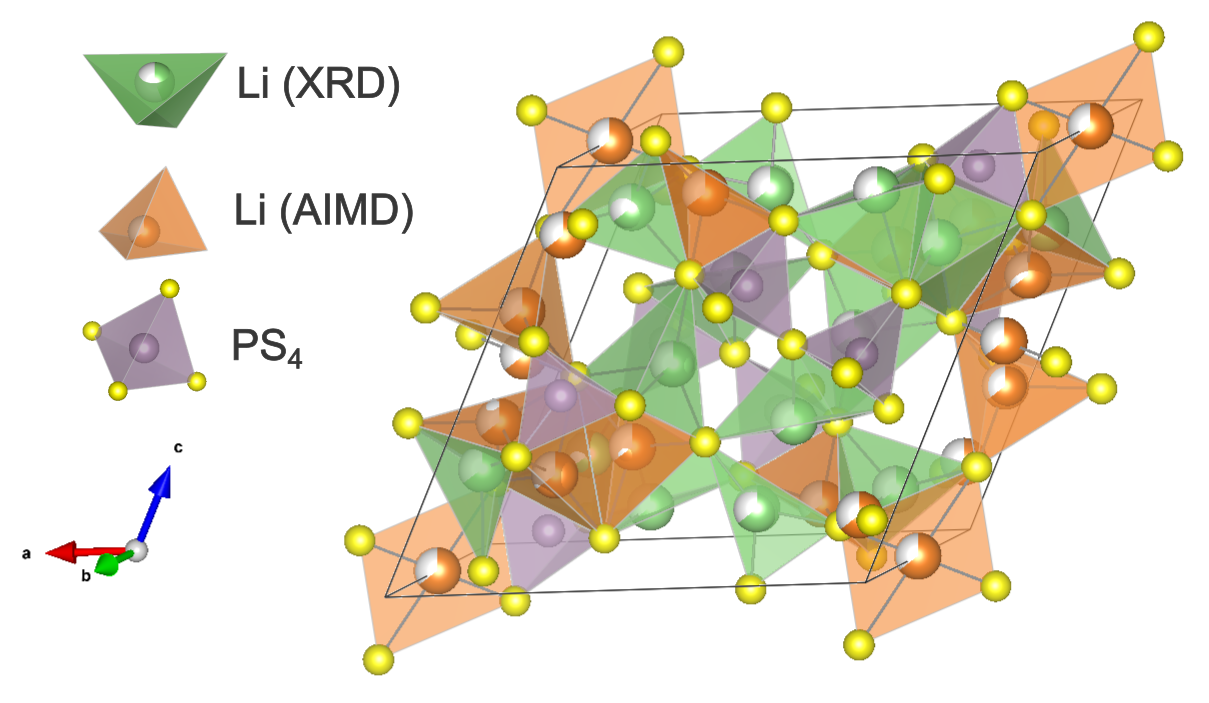}
    \caption{Disordered unit cell of \ch{Li7P3S11} with 7 distinct Li sites identified from XRD (green) and 8 distinct sites identified from AIMD simulations (orange).}
    \label{fig:li7p3s11_prim}
\end{figure}

We train a CE on the previously reported disordered \ch{Li7P3S11} lattice containing 15 distinct Li sites, which are a sum of the sites identified from XRD and AIMD. The unit cell is shown in Figure \ref{fig:li7p3s11_prim}, from which we can observe that the possible Li sites include a range of planar and tetrahedral coordination environments with varying degrees of distortion. Through MC simulated annealing, we uncover a new ground state ordering (SI Figure S4d) that is 21 meV/f.u. (1 meV/atom) more stable than the ground state previously proposed by Chang and coworkers \cite{chang_7311_aimd}.

\begin{figure}
    \centering
     \includegraphics[scale=0.65]{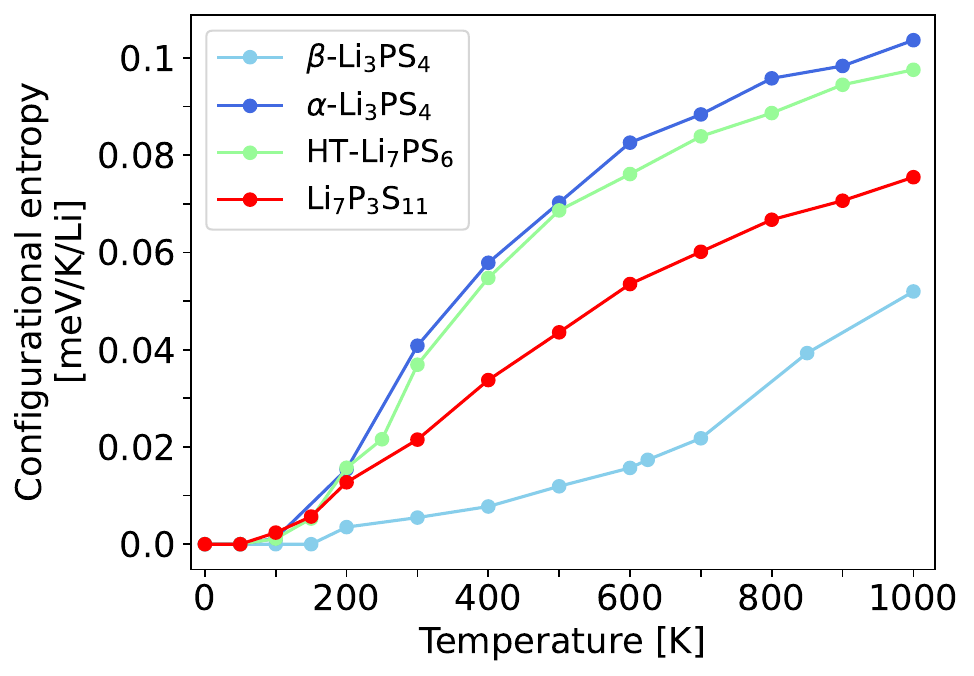}
    \caption{Configurational entropy in the disordered superionic conductors $\alpha$-\ch{Li3PS4}, $\beta$-\ch{Li3PS4}, HT-\ch{Li7PS6}, and \ch{Li7P3S11} phases, normalized per Li atom.}
    \label{fig:config_s_comp}
\end{figure}

Since crystallographic refinements have not reported the existence of configurational disorder in this phase, it is important to quantify the degree of disorder, and compare this with other superionic conductors \cite{yamane_li7p3s11_xrd_2007, onodera_li7p3s11_nd_xrd}. To that end, we calculate the configurational entropy as a function of temperature with MC simulations for \ch{Li7P3S11}, and compare it to that of $\alpha$-\ch{Li3PS4}, $\beta$-\ch{Li3PS4}, and HT-\ch{Li7PS6}, which are plotted in Figure \ref{fig:config_s_comp}. \ch{Li7P3S11} (red) is predicted to contain significant configurational entropy that is greater than $\beta$-\ch{Li3PS4} (light blue), and lower but comparable to $\alpha$-\ch{Li3PS4} (dark blue) and HT-\ch{Li7PS6} (green). This result corroborates the additional Li sites in the disordered Li sublattice identified from AIMD \cite{chang_7311_aimd}. We remark that all superionic conductors in this phase space contain significant configurational entropy that is of the same order of magnitude, indicating a potential correlation between superionic conductivity and configurational entropy.

\section{Discussion}
Through applying a range of first-principles techniques to capture electronic, configurational, and vibrational sources of free energy in the pseudo-binary \ch{Li2S}-\ch{P2S5} system, we recover all experimentally observed polymorph phase transitions in \ch{Li3PS4} and \ch{Li7PS6}, and well-established trends such as the metastability of \ch{Li7P3S11}. An accurate assessment of the configurational entropy required precise information on the possible Li sites in these structures. We find that ND refinements tend to contain more accurate details about Li sites and degree of disorder, compared to XRD refinements. Our first principles calculations show that these details from ND are critical towards predicting physically accurate dynamical stability and thermodynamic behavior.
 
 Vibrational and configurational sources of entropy are shown to be crucial towards describing phase stability trends. Among the \ch{Li3PS4} polymorphs, the superionic conductors $\alpha$ and $\beta$ have distinctly greater vibrational entropy compared to $\gamma$, which has low Li conductivity. We attribute this to the softness of the anion sublattice, as $\alpha$ and $\beta$-\ch{Li3PS4} contain significantly more low-frequency S vibrational modes, mainly corresponding to librations of the \ch{PS4} group. The potential electronic origin of the stiffer anion sublattice in $\gamma$-\ch{Li3PS4} lies in the stronger hybridization of the P 3p and S 3p states near the Fermi level. We postulate that these subtle differences in longer range binding between \ch{PS4} units are the reason why a meta-GGA level of theory is required to predict the correct order of \ch{Li3PS4} polymorph stability, as the SCAN family of density functionals have been shown to be superior at capturing medium-range van der Waals interactions.\cite{sun_scan_2016, julia_rationalize_scan_2019}. These findings can potentially motivate new design principles for novel superionic conductors based on features of the phonon and electronic band structure \cite{muy_sic_phonon}.

Configurational sources of entropy are also essential towards describing phase stability trends. The polymorphic phase transitions involving $\alpha$-\ch{Li3PS4} and HT-\ch{Li7PS6} can only be predicted when accounting for configurational disorder, which in turn requires accurate assessment of possible sites that Li can access. Furthermore, all superionic conductors in this phase space contain a significant amount of configurational entropy. $\beta$-\ch{Li3PS4} has the lowest configurational entropy, and coincidentally its bulk ionic conductivity has been reported to be low (8.9 $10^{-3}$ mS/cm), with only its nanoporous form having high Li conductivity (0.16 mS/cm) \cite{liu_anomalous_2013}. The high temperature $\alpha$ polymorph has considerably greater configurational entropy and room temperature Li conductivity of $\sim$2 mS/cm \cite{alpha_synthesis_2023}. Meanwhile, the $\gamma$ polymorph has the lowest ionic conductivity and contains no configurational disorder. This observation suggests an inherent correlation between fast Li mobility and high configurational entropy.

This trend is observed in many other systems as well. We show that HT-\ch{Li7PS6} has high configurational entropy, comparable to $\alpha$-\ch{Li3PS4}, and it is experimentally shown to have greater Li conductivity than LT-\ch{Li7PS6} \cite{kong_li7ps6_se_nmr}. This trend is not unique to sulfide superionic conductors, as the oxide garnet \ch{Li7La3Zr2O12} (LLZO) has a low-temperature ordered tetragonal phase with low Li conductivity, and a high-temperature disordered superionic conductor with increased cubic symmetry \cite{murugan_llzo_2007}. We observe that superionic conductors tend to be high temperature polymorphs with increased symmetry arising from the configurational disorder. These phases must be entropically stabilized at high temperature, which lends further support that high entropy is favorable towards achieving a superionic conducting state.

\begin{figure}
    \centering
    \begin{subfigure}[b]{.46\linewidth}
        \centering
        \includegraphics[scale=0.65]{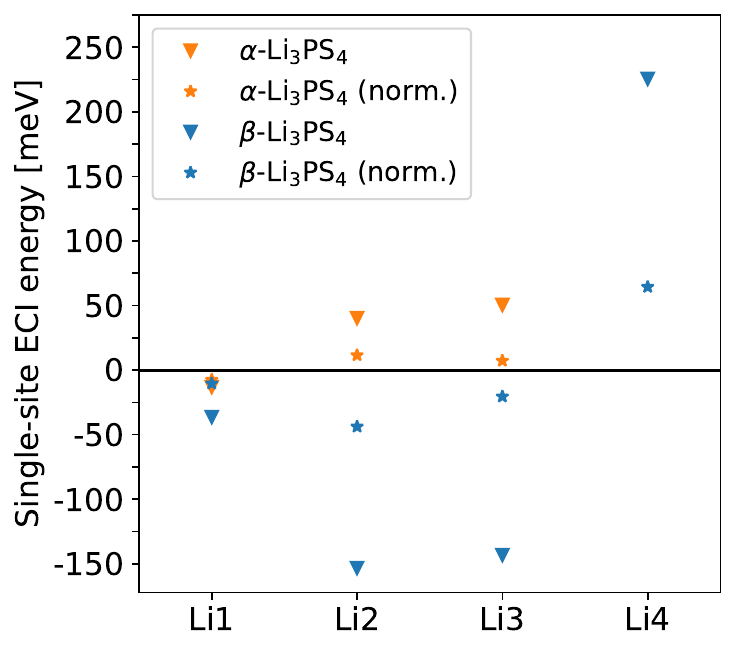}
        \caption{}
        \label{fig:li3ps4_pt_eci}
    \end{subfigure}
    \hspace{0.5cm}
    \begin{subfigure}[b]{.46\linewidth}
        \centering
        \includegraphics[scale=0.63]{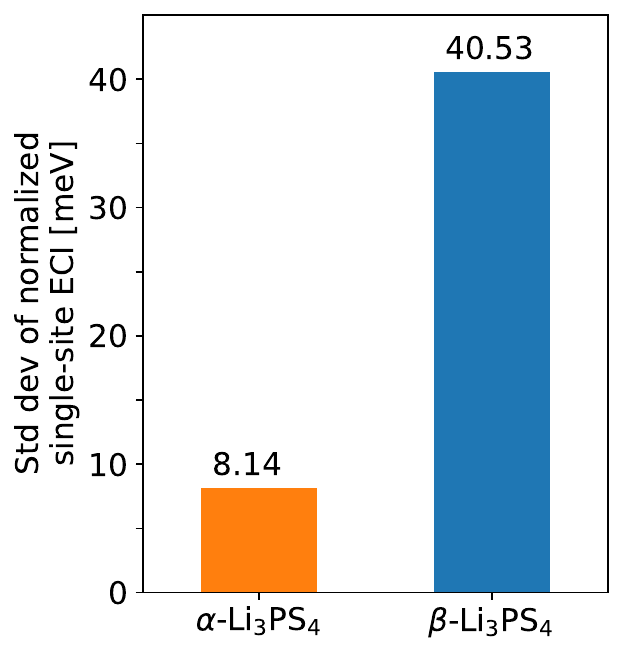}
        \caption{}
        \label{fig:li3ps4_pt_eci_std}
    \end{subfigure}
    \caption{Comparing Li site energies in $\alpha$ and $\beta$-\ch{Li3PS4} polymorphs. a) ECI energies  corresponding to singe-site functions in the CE ($J_{ECI}$, denoted by triangles), and ECI energies normalized by site multiplicity ($\hat{J}_{ECI}$, denoted by stars). In $\beta$, the sites are Li1 8d, Li2 8d, Li3 4c, and Li4 8d. Positive values correspond to an increase in energy from Li occupying a site. In $\alpha$, the sites are Li1 16h, Li2 8e, and Li3 4c. b) Standard deviation of $\hat{J}_{ECI}$.}
\end{figure}

We can rationalize the origin of high configurational entropy by analyzing Li site energies. $\beta$-\ch{Li3PS4} and its higher symmetry $\alpha$-\ch{Li3PS4} polymorph are ideal systems to compare, as they have the same number of Li atoms and Li sites per unit cell. A first order approximation for the Li site energy is the site's effective cluster interaction (ECI) energy ($J_{ECI}$) obtained from the CE using an orthonormal basis, which are plotted in Figure \ref{fig:li3ps4_pt_eci}. It can be shown from the cluster decomposition framework that this is a unique and physical value to describe the energy of Li occupying a particular site \cite{luis_cluster_decomp}. This approximation can be justified by the observation that single-site ECI tend to be much larger in magnitude than the multi-site pair and triplet ECI (SI Figure S5); single-site ECI thus carry most of the weight in the total energy. We also calculate a site energy normalized by its multiplicity ($\hat{J}_{ECI}$), which would provide a better estimate of the energy contribution of the site per unit cell. This is described in Equation \ref{eci_norm}, where $M$ is the multiplicity of a distinct Li site and $N$ is the total number of Li sites per unit cell.
\begin{equation}\label{eci_norm}
    \hat{J}_{ECI} = J_{ECI} \cdot \frac{M}{N}
\end{equation}
In Figure \ref{fig:li3ps4_pt_eci_std}, we plot the standard deviation of $\hat{J}_{ECI}$ in each \ch{Li3PS4} phase, showing that $\alpha$ contains a significantly smaller spread of $\hat{J}_{ECI}$ (8 meV) compared to the $\beta$ polymorph (41 meV). Thus, in $\alpha$ the Li atoms will have a comparable energetic preference for occupying all sites. Many configurations will then have similar energy, which contributes towards its greater configurational entropy. The larger Li site energy spread in $\beta$ means that Li atoms will tend to order by occupying the lowest energy sites and thus have smaller configurational entropy.
 
The potential connection between high Li mobility and configurational entropy suggests a rather obvious design strategy of doping superionic conductor phases to increase configurational disorder. Indeed, there have been many examples where cation or anion doping improves Li conductivity and enables room temperature phase stability. These include adding Si into \ch{Li3PS4} to form Li$_{3.25}$P$_{0.75}$Si$_{0.25}$S$_4$ in the $\beta$-\ch{Li3PS4} structure \cite{zhou_entropically_2019}, adding Cl or other halogen atoms (X) to \ch{Li7PS6} to form \ch{Li6PS5X} in the HT-\ch{Li7PS6} structure \cite{schlenker_li6ps5cl_nd_pdf, kraft_polariz_li6ps5x, adeli_argyrodite_subs_2019}, and doping Al or Ga into \ch{Li7La3Zr2O12} to stabilize its high-temperature cubic structure \cite{karasulu_ga-doped_2020}. We have shown that the disorder arising from only Li and vacancies can generate substantial configurational entropy that can dictate phase stability trends. Introducing doping would generate additional disorder in the non-Li cation or anion sublattices, which should considerably increase configurational entropy and provide greater thermodynamic stability at lower temperatures.

Previous studies have also shown that adding dopant species can alter the Li site energy landscape to facilitate dramatic improvements in ionic conductivity. Zeng and co-workers demonstrated that high-principal element cation doping can boost ionic conductivity by orders of magnitude \cite{zeng_high_entropy_sic_2022}. Through first principles calculations, they showed that distortions to Li environments introduced by dopants can lead to Li site energy levels that are more closely spaced, promoting Li-ion percolation. It is possible that the soft degrees of freedom for libration of the \ch{PS4} units as seen in several polymorphs further generates the distribution of temporary site energies which leads to low energy barrier percolation pathways\cite{forrester_aimd_cat_an_2023}. Similarly, Wang and co-workers found that adding Br into \ch{Li3YCl6} to form Li$_3$YBr$_{1.5}$Cl$_{4.5}$ introduced a larger variety of closely spaced octahedral Li site energy levels, leading to a lower order-disorder transition temperature and increased Li conductivity \cite{wang_doas_2023}. These previous studies highlight that engineering a more uniform Li site energy landscape will facilitate more facile Li-ion migration. We can synthesize this with our finding that smaller variance in Li site energies necessarily leads to greater configurational disorder as well,  which illustrates why the phenomena of superionic conductivity and high configurational entropy should be intrinsically linked. This rationalizes why introducing dopants has been, and should continue to be, an essential design principle for discovering superionic conductors with improved Li conductivity and thermodynamic accessibility.
 
Accurately modeling configurational disorder in each phase could only be achieved after clarifying the details of Li sublattices. We demonstrate that ND refinements of Li sublattices in $\alpha$-\ch{Li3PS4}, $\beta$-\ch{Li3PS4}, and the \ch{Li6PS5Cl} analogue of HT-\ch{Li7PS6} contain critical details such as site splitting and additional sites that XRD could not detect. These additional sites likely lead to more low-energy configurations that are vital for describing thermodynamic behavior. The deficiencies of XRD refinements can be attributed to Li having poor XRD sensitivity due to its small X-ray scattering factor, while the negative scattering length of Li neutrons leads to greater sensitivity in ND. Despite its known limitations, XRD often yields reasonable results in many Li-containing materials, such as Li transition metal oxide cathodes, and remains a standard technique in characterizing Li battery materials. We speculate that the spurious XRD refinements highlighted in this study stem from very high Li mobility, which would smear the detected Li electron density and thus further deteriorate sensitivity. The close agreement between ND and XRD refinements of $\gamma$-\ch{Li3PS4} can then be explained by its low Li conductivity \cite{homma_li3ps4_crystal_2011}. Our discovery of configurational disorder in LT-\ch{Li7PS6} highlights that there may still be additional details about the Li substructures that are yet to be uncovered, which should motivate further experimental and computational studies to refine the Li atomic arrangements.

Although we have predicted the phase stability trends and rationalized them on the basis of configurational and vibrational contributions, our predicted phase transition temperatures tend to underestimate experimentally observed values by about 200 K. The phase stability trends in this system are described on a rather fine energy scale on the order of 10 meV/atom. To highlight the sensitivity of the energy scale, subtle changes such as hypothetically shifting the free energy curve of $\beta$-\ch{Li3PS4} up by 3 meV/atom can already increase the $\gamma$-$\beta$ transition temperature to its experimentally observed window. These small energy differences are easily within the bounds of error in our computational techniques. Specifically, it is known that semi-local density functionals, such as the GGA and r$^2$SCAN functionals used in this study, struggle to capture long-range dipole-induced dipole interactions \cite{sun_scan_2016}, which are likely to be prominent within the S sublattice in these materials. Furthermore, there is remnant self-interaction error in density functional approximations \cite{perdew_self_int_1981}, which can be mitigated by using more computationally expensive hybrid functional techniques \cite{heyd_hse_2004} or many-body treatments of electron correlation \cite{hybertsen_louie_gw_1986}. 
The error from CE configurational energies is compounded onto the DFT error since the CEs are trained on DFT data. On the basis of cross validation (CV) root mean squared error (RMSE), CE energy error ranges from 1 to 5 meV/atom, depending on the phase (SI Figure S6). Furthermore, anharmonic corrections to phonon calculations may yield key differences in the band dispersion and resulting vibrational free energy, as previously demonstrated in the sodium thiophosphate (\ch{Na3PS4}) analogue \cite{gupta_na_anharm_2021}. The facile and long-range nature of Li hopping modes are a potential source of anharmonicity in superionic conductors. Finally, we have treated the configurational and vibrational entropy contributions as independent, as is common in first-principles alloy theory \cite{vdw_vib_config_2002}. A more accurate, but significantly more computationally intensive approach, would be to also include the configurational-dependence of the vibrational entropy, as can be formally done with the CE approach \cite{garbulsky_vib_ce_1994, vdw_ni3al_vib_s_1998}.

\section{Conclusion}
A phase diagram of the pseudo-binary \ch{Li2S}-\ch{P2S5} system has been constructed from first-principles calculations. Well-established experimental trends, such as the phase transitions among \ch{Li3PS4} and \ch{Li7PS6} polymorphs, and the metastability of \ch{Li7P3S11} are recovered. The superionic conductors $\alpha$-\ch{Li3PS4}, $\beta$-\ch{Li3PS4}, HT-\ch{Li7PS6}, and \ch{Li7P3S11} are all predicted to be metastable at 300 K (E$_{\text{hull}}$ = 4, 1, 12, and 4 meV/atom, respectively).  We find that accounting for both vibrational modes and Li configurational degrees of freedom are essential for describing phase stability trends. Physically accurate evaluation of configurational entropy could only be made after clarifying the details of the Li sublattices in the superionic conductors. We demonstrate that these phases all contain significant configurational entropy, which suggests a correlation between high Li configurational entropy and fast Li conduction. Engineering a more uniform Li site energy landscape through doping should thus be an essential design principle for discovering novel superionic conductors with improved thermodynamic stability and Li conductivity at ambient temperature.

\section{Methods}
All electronic structure calculations were performed using the Vienna ab-initio simulation package (VASP) \cite{kresse_efficient_1996}. For the ground state structures of each phase, ionic relaxations were performed with 1e-05 eV convergence in the total energy and 1e-02 eV/$\AA$ in the forces, initially using the generalized gradient approximation (GGA) functional as parameterized by Perdew, Burke, and Ernzerhof (PBE) \cite{perdew_pbe_1996} and projector augmented wave (PAW) potentials \cite{kresse_uspp_paw_1998}. The GGA-converged structure was further relaxed with the meta-GGA r$^2$SCAN functional \cite{furness_r2scan_2020}, with a k-point spacing dependent on the band gap of the PBE calculation, a scheme proposed by Kingsbury and co-workers \cite{kingsbury_r2scan_scan_comp_2022}. The final reported formation energies were obtained from a static calculation with denser k-point spacing of 0.2 $\AA^{-1}$. Applying increased meta-GGA level of theory was essential for capturing physical polymorph phase stability, as $\gamma$-\ch{Li3PS4} and $\beta$-\ch{Li3PS4} had nearly identical electronic formation energies using PBE (SI Table SI). The relaxed structure, total energy, and calculation details for each phase's ground state are attached in the form of a Pymatgen ComputedStructureEntry JSON file in the attached \texttt{lps\char`_final\char`_gs\char`_entries.zip} folder.\cite{ong_pymatgen_2013}

CE construction and MC sampling were performed with the \textsc{smol} Python package \cite{luis_smol_2022}. The primitive structures used to construct the CE for each phase are described in SI Table SII-SV. CEs were trained on superstructures relaxed using the PBE functional only, to limit the computational cost. It has been previously shown that similar schemes of mixing levels of theory can yield physically accurate phase diagrams \cite{kingsbury_mix_gga_scan_2022}. We simultaneously parameterize the CE with an additional electrostatic energy term, which was calculated from the bare Coulomb interaction between idealized Li$^{+}$, P$^{5+}$, and S$^{2-}$ point charges with the Ewald summation method. CE fitting was performed in a piece-wise manner, where the initial fit only trained the point correlation functions and effective dielectric constant ($\epsilon$), using L2 norm penalized linear regression. The residual of the initial fit was used to train the pairs and higher-order effective cluster interactions (ECI) with penalization of the L1 norm. We observed that this method yields improved fit stability and a more physical $\epsilon$, which is attributed to decreased regularization of $\epsilon$ \cite{barroso_luque_ionic_ce_2022}. MC sampling was performed in supercells for each phase in the canonical ensemble, with decreasing temperatures starting at 1000 K. Supercells were constructed to contain at least 200 Li sites and have similar lattice parameters. At least 40000 MC passes were performed at each temperature. To calculate configurational free energy, the average internal energy ($\langle E \rangle$) at each temperature was integrated over inverse thermal energy ($\beta$ $= \frac{1}{\text{kT}}$) (Equation \ref{thermo_int}). 
\begin{equation}\label{thermo_int}
    \beta F_\text{config}(T) = \beta F_\text{config}(T=T_0) + \int_{\beta_0}^\beta \langle E \rangle d\beta
\end{equation}
New ground states were found from simulated annealing, using a similar procedure of canonical MC sampling at decreasing temperature, but with unit cells and smaller supercells.

Harmonic phonon calculations were performed on the ground state of each phase, with the frozen phonon method using Phonopy and VASP \cite{togo_phonopy_2015}. Structures were relaxed with PBE to a stricter convergence criteria of 1e-7 eV in energy and 1e-3 eV/$\AA$ in the forces. Atomic displacements were generated on supercells, which were created such that each lattice parameter is greater than 12 $\AA$ and nearly equal to each other. For \ch{P2S5} only, we calculate the phonon properties using density functional perturbation theory (DFPT) as implemented in VASP \cite{gajdos_vasp_dfpt_2006}, because we observed that the frozen phonon method yielded many imaginary modes, which we attribute to a strongly anharmonic potential energy surface. Non-analytical correction to modes near the $\Gamma$ wave vector was performed by incorporating the dielectric properties, to account for longitudinal optical and transverse optical (LO-TO) mode splitting in polar ionic materials in the long wavelength limit \cite{gonze_dfpt_1997}. Dielectric permittivity and Born effective charge tensors were computed with DFPT \cite{gajdos_vasp_dfpt_2006}, using a denser reciprocal space discretization of 0.125 $\AA^{-1}$ to ensure convergence of dielectric properties \cite{petousis_high_thru_dielectric_2017}. The vibrational free energies and phonon total density of states for each phase (if not already shown in the previous sections) are plotted in SI Figures S8 and S9, respectively.

Using the electronic, configurational, and vibrational free energies, the formation free energies and resulting phase diagrams were computed with Pymatgen \cite{ong_pymatgen_2013}. The formation free energies of phases relative to the \ch{Li2S} and \ch{P2S5} end points are shown in SI Figure S7.

\bibliography{lps_refs_2_edited.bib}

\section{Acknowledgements}
The authors would like to thank Prof. Kristin Persson, Prof. Geoffrey Hautier, and Sunny Gupta for useful insights on phonon calculations. This work was supported by the Assistant Secretary of Energy Efficiency and Renewable Energy, Vehicle Technologies Office of the US Department of Energy (DOE), under contract no. DE-AC02-05CH11231 under the Advanced Battery Materials Research (BMR) Program. This research used resources of the National Energy Research Scientific Computing Center (NERSC), a U.S. Department of Energy Office of Science User Facility operated under contract no. DE-AC0205CH11231, and the Extreme Science and Engineering Discovery Environment (XSEDE), which is supported by the National Science Foundation grant number ACI1053575.
\end{document}